\documentclass[a4paper,fleqn,usenatbib]{mnras}

\usepackage[T1]{fontenc}
\usepackage{ae,aecompl}

\usepackage{graphicx}
\usepackage{bm}
\usepackage{amssymb}
\usepackage{amsmath}
\usepackage{amsfonts}
\usepackage{dcolumn}
\usepackage{natbib}
\usepackage{color}
\usepackage{calc}

\usepackage[normalem]{ulem}

\newcommand{\raiseentry}[1]{\smash{\raise 0.7 em \hbox{#1}}}

\newcommand{\lowentry}[1]{\smash{\lower 1.5 ex \hbox{#1}}}

\def\apj{Astrophys. J.}
\def\apjl{Astrophys. J. Lett.}
\def\apjs{Astrophys. J. Supp. Ser. }

\def\aap{Astron. Astrophys. }

\def\mnras{Mon. Not. Roy. Astron. Soc. }

\newenvironment{equationarray*}
{\arraycolsep 0.14 em
\begin{eqnarray*}}
{\end{eqnarray*}}

\title{Shock-Turbulence Interaction in Core-Collapse Supernovae}

\author[E. Abdikamalov et al.]{
Ernazar Abdikamalov$^{1}$,\thanks{ernazar.abdikamalov@nu.edu.kz} 
Azamat Zhaksylykov$^{1}$,
David Radice$^{2}$,
and Shapagat Berdibek$^{1}$
\\
$^{1}$Department of Physics, School of Science and Technology, Nazarbayev University, The Republic of Kazakhstan\\
$^{2}$TAPIR, MC 350-17, California Institute of Technology, 1200 E California Blvd., Pasadena, CA 91125, USA
}

\date{May 29, 2016}
\pubyear{2016}

\begin{document}
\label{firstpage}
\pagerange{\pageref{firstpage}--\pageref{lastpage}}
\maketitle

\begin{abstract}
Nuclear shell burning in the final stages of the lives of massive stars is accompanied by strong turbulent convection. The resulting fluctuations aid supernova explosion by amplifying the non-radial flow in the post-shock region. In this work, we investigate the physical mechanism behind this amplification using a linear perturbation theory. We model the shock wave as a one-dimensional planar discontinuity and consider its interaction with vorticity and entropy perturbations in the upstream flow. We find that, as the perturbations cross the shock, their total turbulent kinetic energy is amplified by a factor of $\sim\!2$, while the average linear size of turbulent eddies decreases by about the same factor. These values are not sensitive to the parameters of the upstream turbulence and the nuclear dissociation efficiency at the shock. Finally, we discuss the implication of our results for the supernova explosion mechanism. We show that the upstream perturbations can decrease the critical neutrino luminosity for producing explosion by several percent. 
\end{abstract}



\section{Introduction}
\label{section:introduction}

Massive stars undergo vigorous convective shell burning at the end of their lives \cite[e.g.,][]{arnett:09,
takahashi:14, couch:15b, mueller:16b, chatzopoulos:16}.  The associated
non-radial dynamics and the deviations from spherical symmetry
can grow further during collapse \citep{lai:00n,takahashi:14}.  Recent
works by \citet{couch:13d, couch:15a} and \citet{mueller:15} demonstrate
that such asphericities facilitate supernova explosion. According to
\citet{couch:15a, mueller:15}, this is a result of increased turbulent
activity in the post-shock region driven by the passage of the upstream
fluctuations through the shock. The non-radial dynamics in the
post-shock region is an important factor that aids the expansion of the
supernova shock \citep[e.g.,][]{herant:95, bhf:95, janka:96, blondin:03,
foglizzo:06, foglizzo:07, hanke:12, hanke:13, janka:12b, dolence:13,
murphy:13, burrows:13a, takiwaki:14, ott:13a, abdikamalov:15,
radice:15a, melson:15a, melson:15b, lentz:15, fernandez:15a,
foglizzo:15, cardall:15, radice:16a, bruenn:16, janka:16, roberts:16n}.

In this work, we investigate the physics of the interaction of the upstream turbulence with the supernova shock and its effect on the post-shock flow using a linear perturbation theory commonly known as the linear interaction approximation (LIA) theory. The LIA, which we extend to include the nuclear dissociation at the shock, is a powerful tool originally developed in the 1950s by \cite{ribner:53}, \cite{moore:54}, and \cite{chang:57}, followed by other works \citep[e.g.,][]{ribner:54, chang:57, mckenzie:68, jackson:90, mahesh:96, mahesh:97, duck:97, fabre:01, wouchuk:09, huete:11, huete:12}.

In the LIA, the shock is modeled as a planar discontinuity with
no intrinsic scale and the flow is decomposed into the mean and
fluctuating parts. Both components can be specified arbitrarily in the
upstream flow. Once the upstream field is specified, the downstream
field can be fully determined using the Rankine-Hugoniot jump conditions
at the shock \citep[e.g.,][]{sagaut:08}. The LIA is valid in
the regime of sufficiently small fluctuations such that the mean flow
satisfies the usual jump conditions, while the turbulent fluctuations
satisfy the linearized jump conditions. Numerical simulations by
\cite{lee:93b} suggest that this approximation is valid when
\begin{equation}
\label{eq:lia}
{\cal M}'^2\lesssim 0.1 ({\cal M}^2_1-1),
\end{equation}
where ${\cal M}'$ and ${\cal M}_1$ are the Mach number of upstream turbulence and mean flow, respectively \citep[see also][]{ryu:14}. In massive star shell convection, ${\cal M}' \sim 0.1$ \citep[e.g.,][]{mueller:16b}, which at most can increase by a factor of several during contraction (more precise calculation of this is given below in Section \ref{sec:implications}). Since ${\cal M}_1 \gtrsim 5$ in core-collapse supernovae (CCSNe), condition (\ref{eq:lia}) is well satisfied and we expect the LIA to be an excellent approximation for studying the interaction of CCSN shocks with progenitor asphericities. 

\section{The Linear Interaction Approximation}
\label{sec:lia}

The LIA employs the \citet{kovasznay:53} decomposition of the
fluctuating field, according to which any small fluctuations in a
turbulent flow can be decomposed individual Fourier modes that are
characterized by their type, wavenumber, and frequency. There are three
types of modes: vorticity, entropy, and acoustic modes. The vorticity
mode is a solenoidal velocity field that is advected with the mean flow.
It has no pressure or density fluctuations. The entropy mode is also
advected with the flow and it represents density and temperature
fluctuations with no associated pressure or velocity variations. The
acoustic mode represents sound waves that travel relative to the mean
flow. It has isentropic pressure and density fluctuations and
irrotational velocity field. All Kovasznay modes evolve independently in
the limit of weak fluctuations and the interaction of each mode with the
shock wave can be studied independently. Integration over all individual
modes yields the full statistics of the turbulent flow
\citep[e.g.,][]{sagaut:08}. 

We assume that the shock wave is a planar discontinuity and we choose our $x$-axis ($y$-axis) to be perpendicular (parallel) to the shock front. The average shock position is assumed to be at $x=0$ and the mean flow is in the positive $x$ direction. The quantities $U$,  $\bar{\rho}$, $\bar{p}$, $\bar{T}$, and $\mathcal{M}$ represent the mean velocity, density, pressure, temperature, and Mach number. We choose the values of these parameters to approximate the CCSN shock by requiring vanishing Bernoulli parameter for the upstream flow, as described in Appendix \ref{app:shock}. We employ a gamma-law equation of state with $\gamma=4/3$. The quantities $u'$, $\upsilon'$ $\rho'$, $p'$, $T'$ denote the perturbation in the $x$- and $y$-components of velocity, density, pressure, and temperature, respectively. Hereafter, subscripts 1 and 2 will denote the upstream and downstream states. The upstream vorticity mode is modeled via a planar shear wave with wavenumber $(m\kappa,l\kappa)$ and angular frequency $\kappa mU_1$:
\begin{eqnarray}
\label{eq:u1}
\frac{u_1'}{U_1} &=& lA_\upsilon e^{i\kappa(mx+ly-U_1mt)}, \\
\frac{\upsilon_1'}{U_1} &=& -mA_\upsilon e^{i\kappa(mx+ly-U_1mt)},
\label{eq:v1}
\end{eqnarray}
while the upstream entropy mode is given by another planar sinusoidal wave with the same wavenumber and frequency,
\begin{eqnarray}
\frac{\rho_1'}{\bar{\rho}_1} &=& A_e e^{i\kappa(mx+ly-U_1mt)}, \\
\label{eq:rho1}
\frac{T_1'}{\overline{T}_1} &=& -\frac{\rho'}{\bar{\rho}_1},
\label{eq:T1}
\end{eqnarray}
where $m=\cos\psi_1$ and $l=\sin\psi_1$ and $\psi_1$ is the angle
between the $x$-axis and the direction of propagation of the incident
perturbation. $A_\upsilon$ and $A_e$ are the amplitudes of the incident
vorticity and entropy waves, respectively. In the present work, we
ignore acoustic waves in the pre-shock region, which corresponds to the assumption of zero
pressure fluctuations in the upstream flow ($p_1' = 0$). The effect of the upstream acoustic
component will studied in our future work. 

\begin{figure}
 \includegraphics[angle=0,width=1\columnwidth,clip=false]{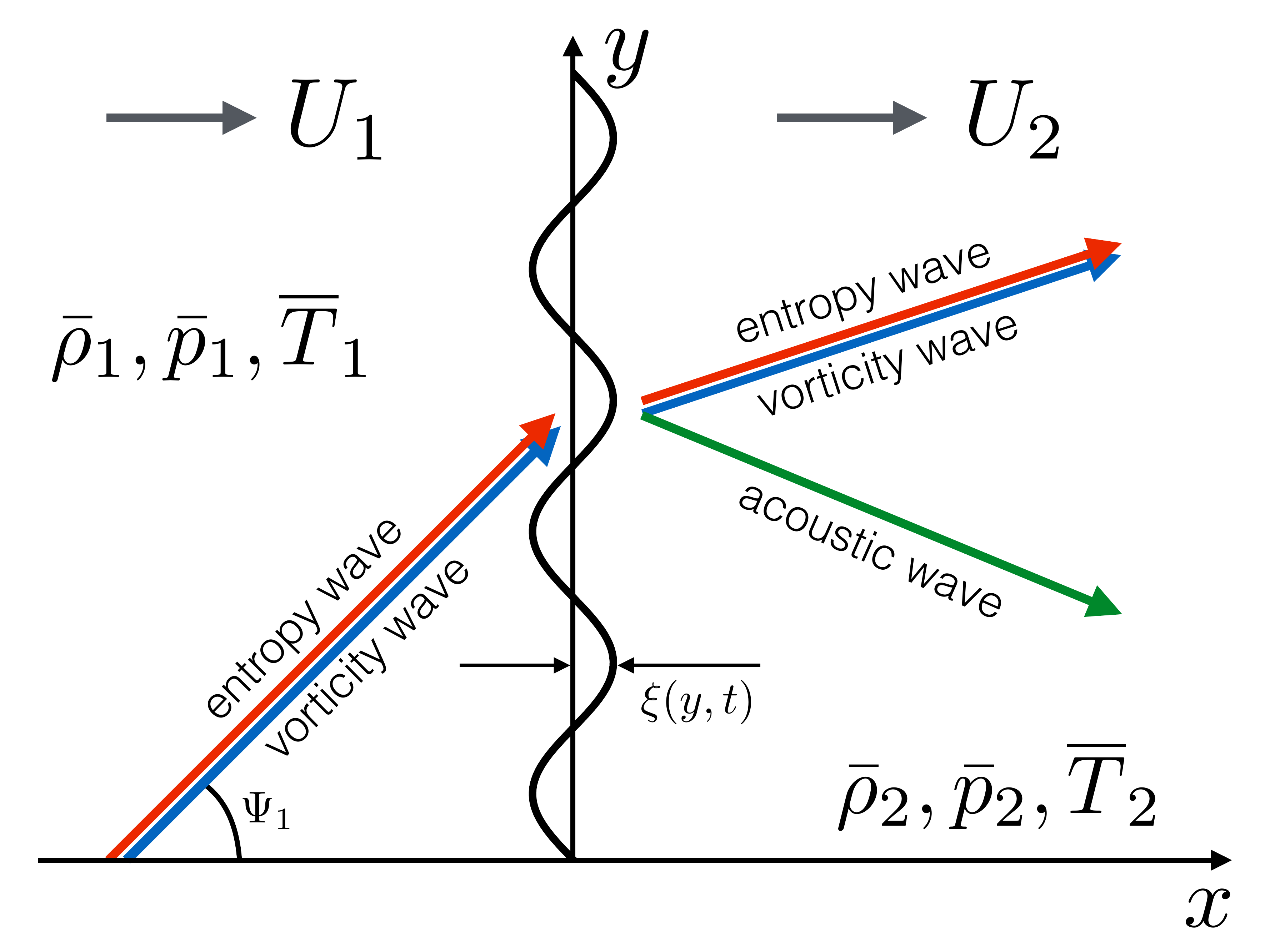}
  \caption{Schematic representation of the interaction of an entropy and/or vorticity waves with a shock wave in the context of the LIA formalism. The average position of the shock is aligned with the $y$-axis and the mean flow is in the positive $x$ direction. The upstream mean flow is characterized by velocity $U_1$, density $\bar{\rho}_1$, pressure $\bar p_1$, and temperature $\overline T_1$, while the corresponds downstream quantities are $U_2$, $\bar{\rho}_2$, $\bar p_2$, and $\overline T_2$. When vorticity and/or entropy waves of form (\ref{eq:u1})-(\ref{eq:T1}) hit a shock wave, the latter responds by changing its position and shape. In the framework of LIA, for such perturbations, the shock surface deforms into a sinusoidal planar wave propagating in the $y$-direction described by formula (\ref{eq:shock_lia0}). The downstream perturbation field consists of entropy, vorticity, and acoustic waves given by Eqs.~(\ref{eq:u2})-(\ref{eq:T2}). 
  \label{fig:shocklia}}
\end{figure}

When vorticity and/or entropy waves hit a shock wave, the former responds by changing its position and shape. In the framework of the LIA, for a perturbation of form (\ref{eq:u1})-(\ref{eq:T1}), the shock surface deforms into a shape of a sinusoidal wave propagating in the $y$-direction:
\begin{equation}
\label{eq:shock_lia0}
\xi(y,t) = -\frac{L}{i\kappa m} e^{i\kappa (ly-U_1mt)}, \\
\end{equation}
where $\xi(y,t)$ is the $x$-coordinate of the shock position at time $t$ and ordinate $y$ and $L$ is a quantity that characterizes the amplitude of the shock oscillations (cf. Fig.~\ref{fig:shocklia}).

The interaction of the vorticity and entropy waves with the shock generates a downstream perturbation field consisting of vorticity, entropy, and acoustic waves given by~\citep{mahesh:96,mahesh:97} 
\begin{eqnarray}
\label{eq:u2}
&&\mkern-60mu\frac{u_2'}{U_1} = F e^{i\tilde{k}x}e^{i\kappa(ly-U_1mt)} + Ge^{ik(\mathcal{C}mx+ly-U_1 mt)}, \\ 
\label{eq:v2}
&&\mkern-60mu\frac{\upsilon_2'}{U_1} = H e^{i\tilde{k}x}e^{i\kappa(ly-U_1mt)} + Ie^{i\kappa(\mathcal{C}mx+ly-U_1 mt)}, \\ 
\label{eq:p2}
&&\mkern-60mu\frac{p_2'}{\bar{p}_2} = K e^{i\tilde{k}x}e^{ik(ly-U_1mt)} \\ 
&&\mkern-60mu\frac{\rho_2'}{\bar{\rho}_1} = \frac{K}{\gamma} e^{i\tilde{\kappa}x}e^{ik(ly-U_1mt)} + Qe^{i\kappa(\mathcal{C}mx+ly-U_1 mt)}, \\ 
\label{eq:T2}
&&\mkern-60mu\frac{T_2'}{\overline{T}_1} = \frac{(\gamma-1)K}{\gamma} e^{i\tilde{\kappa}x}e^{i\kappa(ly-U_1mt)} -Qe^{i\kappa(\mathcal{C}mx+ly-U_1mt)}.
\end{eqnarray}
The schematic representation of this process is depicted in
Fig.~\ref{fig:shocklia}. Note that these waves have the same angular
frequencies and $y$-components of wavenumbers as those of the upstream
waves (\ref{eq:u1})-(\ref{eq:T1}). The coefficients $F$, $H$, and $K$ are the amplitudes of the
acoustic component, while coefficients $G$, $I$, and $Q$ are associated
with the entropy and vorticity components. The former two components
have the same wavenumber vector $(m{\cal C}
\kappa,l\kappa)$ and angular frequency $\kappa mU_1$, for which reason
they are often referred to as entropy-vorticity waves. The acoustic
component has the same angular frequency but different wavenumber
$(\tilde{\kappa},l\kappa)$, where $\tilde{\kappa}$ is calculated in
Appendix \ref{app:lia}. The parameter $\mathcal{C}$ is the compression
factor at the shock, $\mathcal{C}=\bar{\rho}_2/\bar{\rho}_1=U_1/U_2$,
which can be obtained from the Rankine-Hugoniot condition as (cf.
Appendix \ref{app:shock}):
\begin{equation}
\label{eq:c}
{\mathcal C} = \frac{\gamma+1}{\gamma+\frac{1}{M_1^2} - \sqrt{\left(1-\frac{1}{M_1^2}\right)^2+(\gamma+1)\frac{(\gamma-1){\cal M}_1+2}{{\cal M}_1^2} \bar \epsilon}}.
\end{equation}
Here, $\bar \epsilon$ is the dimensionless nuclear dissociation parameter, which characterizes nuclear dissociation energy, as explained in Appendix \ref{app:shock}. It typically ranges from $0$, which represents the limit corresponding to zero nuclear dissociation, to $0.4$, which represents strong nuclear dissociation. We adopt $\bar\epsilon=0.2$ and ${\cal M}_1=5$ as our fiducial values. 

In order to obtain the coefficients $F$, $G$, $H$, $I$, $K$, $Q$, $L$,
we first expand the Rankine-Hugoniot conditions to the first order in
amplitudes of incoming perturbations \citep{ribner:53, chang:57,
mahesh:97}. The solution of the resulting equations yield the
coefficients $F$, $G$, $H$, $I$, $K$, $Q$, $L$, as we demonstrate in
Appendix~\ref{app:lia}. 

The downstream acoustic component depends strongly on the
incidence angle $\psi_1$. If $\psi_1$ is smaller than the critical angle
\begin{equation}
\label{eq:psic}
\psi_c = \cot^{-1} \sqrt{\frac{c_{\mathrm{s},2}^2}{U_1^2} - \frac{U_2^2}{U_1^2}}, 
\end{equation}
where $c_{\mathrm{s},2}$ is the downstream speed of sound, then $\tilde{\kappa}$ is real and the sound waves represent freely propagating planar sine waves. On the other hand, if $\psi_c<\psi_1$, $\tilde{\kappa}$ is complex and the solution represents an exponentially-damping planar sine wave \citep{mahesh:96,mahesh:97}. 

A detailed derivation of the LIA equations, including angle $\psi_c$
(\ref{eq:psic}) and wavenumber $\tilde \kappa$ in
Eqs.~(\ref{eq:u2})-(\ref{eq:T2}), is presented in
Appendix~\ref{app:lia}. 

\section{Results}
\label{sec:results}

\begin{figure}
 \includegraphics[angle=0,width=1\columnwidth,clip=false]{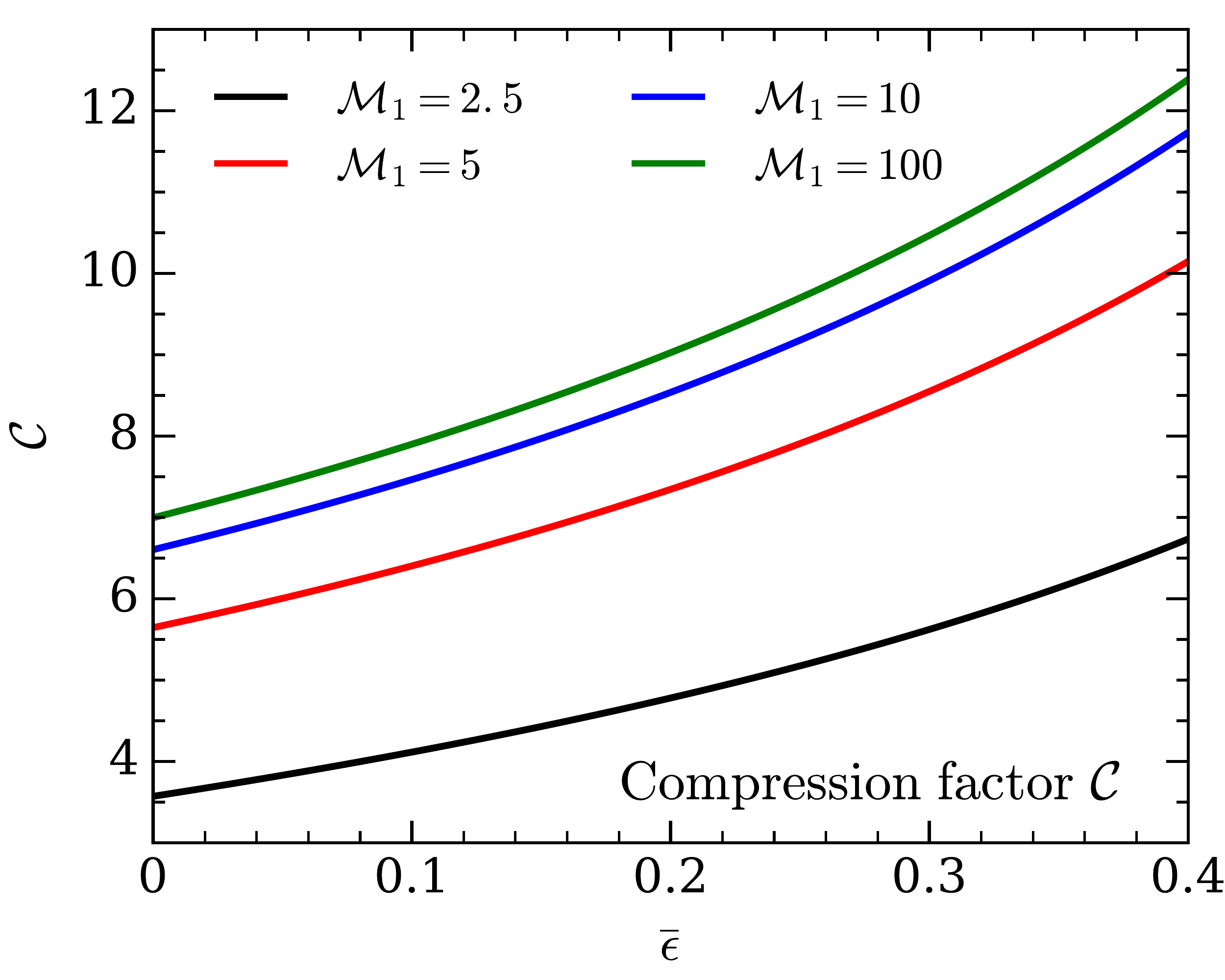}
  \caption{The compression factor $\mathcal{C}$ at the shock as a function of nuclear dissociation parameter $\bar\epsilon$ for four values of upstream Mach number ${\cal M}_1$: $2.5$, $5$, $10$, and $100$. In all of these cases, the compression factor $\mathcal{C}$ increases with $\bar\epsilon$, signifying that the nuclear dissociation leads to stronger compression. 
  \label{fig:c}}
\end{figure}

The key quantity affecting the evolution of the flow through a shock
wave is the compression factor $\mathcal{C}$. Fig.~\ref{fig:c} shows
$\mathcal{C}$ as a function of the nuclear dissociation parameter $\bar
\epsilon$ for four values of upstream Mach number ${\cal M}_1$: $2.5$,
$5$, $10$, and $100$. For all of these values, the compression factor
$\mathcal{C}$ grows with increasing $\bar\epsilon$, meaning that the
nuclear dissociation leads to stronger compression.
Note that the values of the compression factor
$\cal C$ are very close to each other for ${\cal M}_1=5$, $10$, and
$100$. This is a generic property of shock waves, in which the
compression factor depends on ${\cal M}_1$ very weakly when ${\cal
M}_1\gtrsim 5$.

In the following, we present our results in two parts. In the first part (Section~\ref{sec:wave}), we discuss interaction of a shock wave with individual incident waves and explore how it depends on shock and perturbation parameters. In the second part (Section~\ref{sec:turbulence}), we investigate the interaction of a shock wave with incident turbulence fields, which we model as sets of random entropy and vorticity waves. 

\subsection{Interaction with a Single Wave}
\label{sec:wave}

\begin{figure}
 \includegraphics[angle=0,width=1\columnwidth,  clip=false]{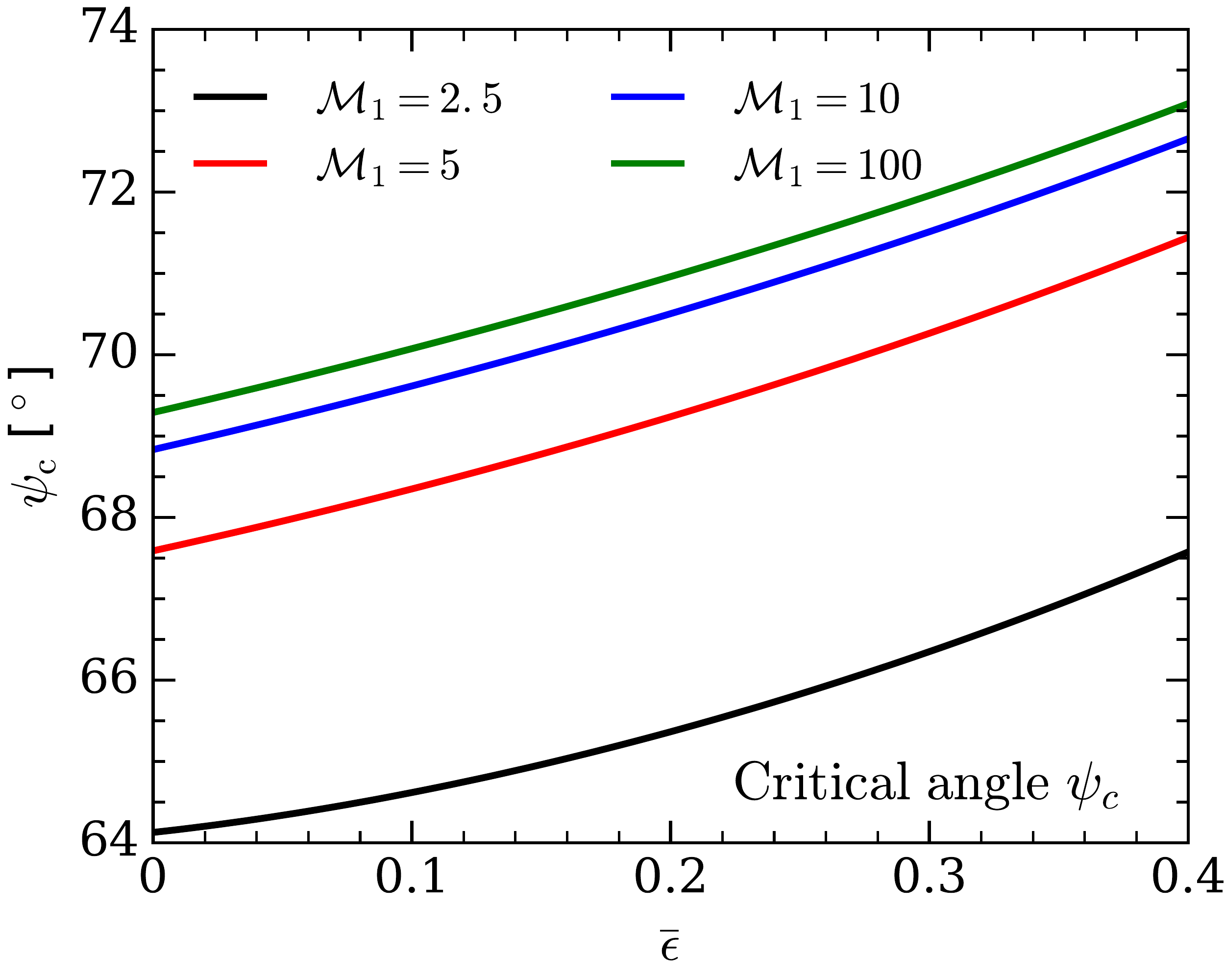}
  \caption{The critical angle $\psi_\mathrm{c}$ as a function of nuclear dissociation parameter $\bar\epsilon$ for four values of upstream Mach number ${\cal M}_1$: $2.5$, $5$, $10$, and $100$.  
  \label{fig:psic}}
\end{figure}

Figure~\ref{fig:psic} shows the values of the critical angle $\psi_\mathrm{c}$ as a function of nuclear dissociation parameter $\bar\epsilon$ for four values of upstream Mach number ${\cal M}_1$: $2.5$, $5$, $10$, and $100$. Recall that the critical angle $\psi_c$ separates two regions of the solution: propagative ($\psi_1 < \psi_c$) and non-propagative ($\psi_1 > \psi_c$). The first is characterized by acoustic waves in the post-shock flow, while in the second sound waves do not propagate. In all cases, $\psi_\mathrm{c}$ increases with $\bar\epsilon$. However, this increase is rather modest. For example, for ${\cal M}_1=5$, $\psi_\mathrm{c}$ increases from $67.6^\circ$ to only $71.5^\circ$ as $\bar\epsilon$ increases from $0$ to $0.4$. These values do not change much with ${\cal M}_1$ after $\mathcal{M}_1=5$. This is a simple reflection of the above-mentioned fact that the compression factor does not depend strongly on the upstream Mach number for ${\cal M}_1 \gtrsim 5$.

For an incident vorticity wave of form given by Eqs.~(\ref{eq:u1})-(\ref{eq:v1}), the velocity field is $u_1' \, \propto \, \sin \psi_1$ and $\upsilon_1' \, \propto \, \cos \psi_1$. If the perturbation wavenumber vector $\vec{k}$ is perpendicular to the shock wave ($\psi_1 = 0$), the $x$-component of the fluctuation field is zero. When such a field hits the shock, the solution is trivial: the shock wave is not affected and the the velocity passes through the shock without any modifications, i.e., $\upsilon_2'=\upsilon_1'$ and $u_2'=u_1'=0$. The only property that changes is the $x$-component of the wavenumber: it increases by a factor of $\cal C$. Correspondingly, the wavelength of the wave decrease by the same factor. 

The situation is drastically different when $\psi_1 > 0$. The perturbation velocity field now has non-zero $x$-component, which forces the shock surface to oscillate according to Eq.~(\ref{eq:shock_lia0}). Because of this, the downstream field now consists of not only vorticity waves, but also of entropy and acoustic waves, as described by Eqs.~(\ref{eq:u2})-(\ref{eq:T2}). Both entropy and vorticity waves in the post-shock region have the same wavenumber vector $({\cal C} m\kappa, l\kappa)$. Thus, the magnitude of the wavenumber vector increases by factor
\begin{equation}
\label{eq:kappa_ratio}
\frac{\kappa_2}{\kappa_1} = \sqrt{{\cal C}^2m^2+l^2}.
\end{equation}
as the wave crosses the shock. Accordingly, the wavelength of the mode decreases by the same factor across the shock. 

\begin{figure}
 \includegraphics[angle=0,width=1\columnwidth, clip=false]{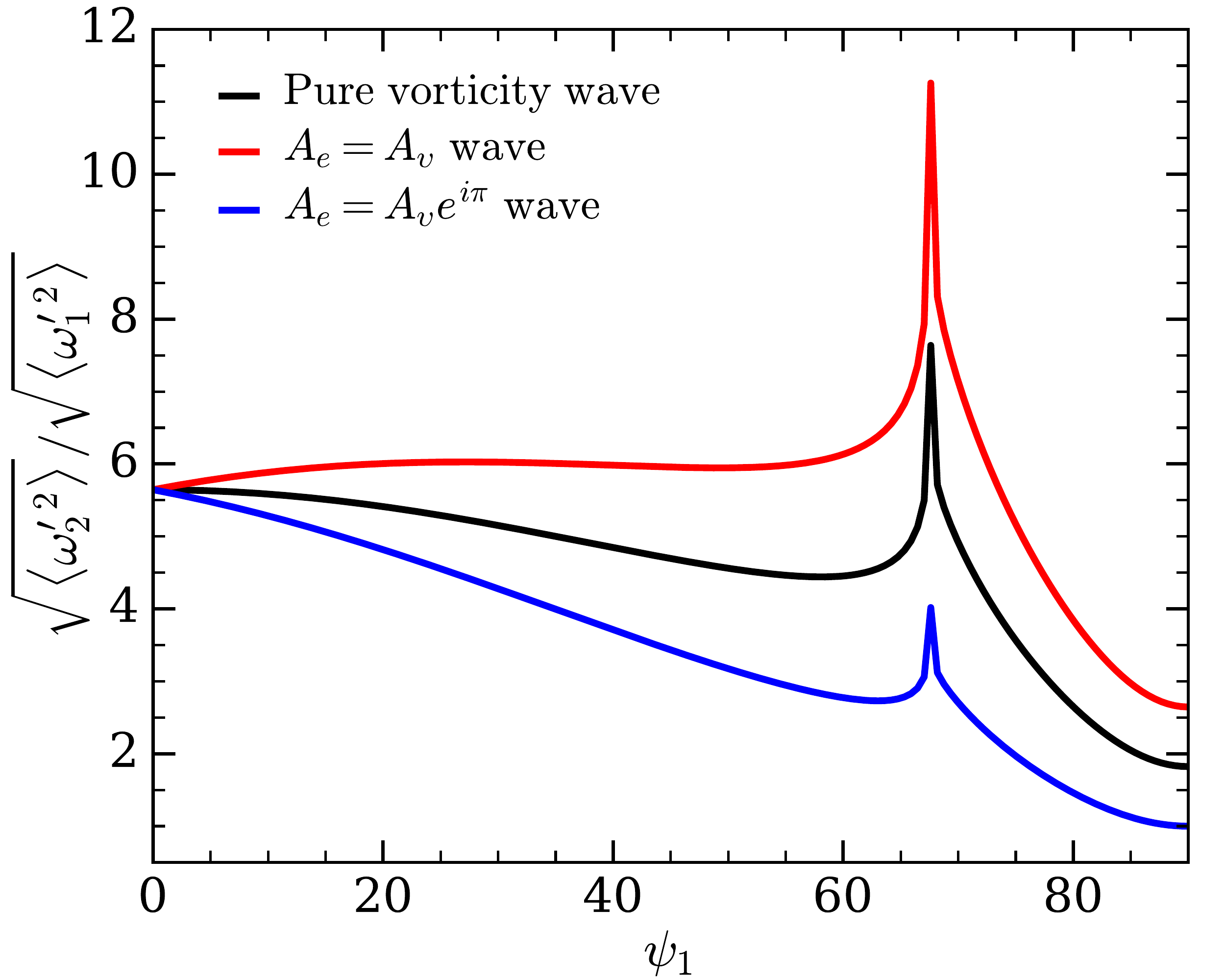}
  \caption{The amplification of vorticity across the shock as a function of angle $\psi$ for incident vorticity wave, in-phase vorticity-entropy wave (i.e., $A_e=A_\upsilon$), and out-of-phase vorticity-entropy wave (i.e., $A_e=A_\upsilon e^{i\pi}$). The upstream Mach number is ${\cal M}_1=5$. When the vorticity and entropy waves are in phase, we get strongest amplification. When they are out of phase, we get the weakest amplification. 
  \label{fig:w2w1_phi}}
\end{figure}

Figure~\ref{fig:w2w1_phi} shows the ratio of averaged pre-shock and post-shock vorticities $\sqrt{\langle\omega_2'^2\rangle}/\sqrt{\langle\omega_1'^2\rangle}$ as a function of angle $\psi_1$ for $\mathcal{M}_1=5$ and $\bar\epsilon=0$\footnote{Note that since the flow is restricted to $x$-$y$ plane, vorticity has only the $z$-component:  
\begin{equation}
\omega_1' = \partial_x \upsilon_1' - \partial_y u_1'=-ikA_\upsilon e^{ik(mx+ly-U_1mt)}.
\end{equation}
As we show below, this is not a restriction because a general 3D problem can be expressed in terms of a 2D LIA problem.}. Here, brackets $\langle\rangle$ mean averaging over time $t$ and the $y$-coordinate. The solid black line represents the case of incident vorticity wave. The case of incident entropy and vorticity waves of the same amplitude and phase (i.e., $A_e=A_\upsilon$) is shown with red line, while the same with $180^\circ$ phase difference (i.e., $A_e=A_\upsilon e^{i\pi}$) is represented by the blue line. As we can see, when the vorticity and entropy waves are in phase, we get a significantly stronger amplification. When they are out of phase, we get the weakest amplification. Finally, in the case of pure vorticity incident wave, the amplification is roughly the average of these two regimes. For example, for $\psi_1=60^\circ$, we get $\sqrt{\langle\omega_2'^2\rangle}/\sqrt{\langle\omega_1'^2\rangle}$ of $6.14$, $2.77$, and $4.46$ in these three cases. The spike in $\sqrt{\langle\omega_2'^2\rangle}/\sqrt{\langle\omega_1'^2\rangle}$ around $\psi_1\simeq 69^\circ$ corresponds to the critical angle $\psi_1=\psi_\mathrm{c}$. 

In order to explain the behavior of the vorticity fluctuations $\omega'$, \citet{mahesh:96,mahesh:97} developed a simple model, which we present here for completeness. Linearizing the Euler equations about the mean flow and neglecting the incident pressure perturbations, we get the following equation for the vorticity fluctuations $\omega'$ \citep{mahesh:96}
\begin{equation}
\label{eq:omega2}
\omega_t'+U\omega_x'=-\omega'U_x - \frac{\rho_y'}{\bar{\rho}^2}\bar{p}_x,
\end{equation}
where subscripts $t$ and $x$ mean partial derivatives with respect to these variables. The first term on the right-hand side of this equation ($-\omega'U_x$) represents the effect of the bulk compression. Since velocity drops across the shock, it amplifies the vorticity. The second term ($-\frac{\rho_y'}{\bar{\rho}^2}\bar{p}_x$) represents the baroclinic processes, which produce vorticity even from pure entropy perturbations. It can either amplify or weaken the effect of bulk compression depending on the relative phase between the vorticity and entropy waves. For incident vorticity and entropy waves of form (\ref{eq:u1})-(\ref{eq:T1}), Eq.~(\ref{eq:omega2}) reduces to 
\begin{equation}
\label{eq:omega3}
-\omega'U_x - \frac{\rho_y'}{\bar{\rho}^2}\bar{p}_x \sim A_\upsilon
UU_x - A_e l \frac{p_x'}{\bar{\rho}}
\end{equation}
Since $U_x<0$ and $\bar{p}_x>0$ at the shock, the two sources of vorticity have the same sign if $A_e$ and $A_\upsilon$ have the same sign. In this case, the entropy wave enhances the amplification of vorticity across the shock. If the signs are opposite, the entropy wave weakens the vorticity amplification.

Using Eq.~(\ref{eq:omega3}), we can derive an approximate expression for the value of the downstream vorticity in terms of its upstream value (see Section 3.6 \cite{mahesh:96} for full derivation):
\begin{equation}
\label{eq:omega2_scaling}
\omega_2' \sim {\mathcal C} \omega_1' + \frac{ik\sin \psi_1}{3}A_e U_1 
\frac{1-{\mathcal C}^3}{{\cal C}^2}
\end{equation}
where ${\mathcal C}$ is the compression factor (\ref{eq:c}). This suggests that the incident vorticity wave amplifies by a factor of $\mathcal{C}$ due to shock-compression, while the vorticity created by the incident entropy wave is $\propto kA_e\sin \psi_1 (1-{\mathcal C}^3)/{\mathcal C}^2$.

A word of caution is in order here. The effects due to the change of shock position and shape are absent in Eq.~(\ref{eq:omega2_scaling}) and thus it has a limited quantitative accuracy. Nevertheless, as we will see below, it describes well some key qualitative aspects of our results.

\begin{figure}
 \includegraphics[angle=0,width=1\columnwidth, clip=false]{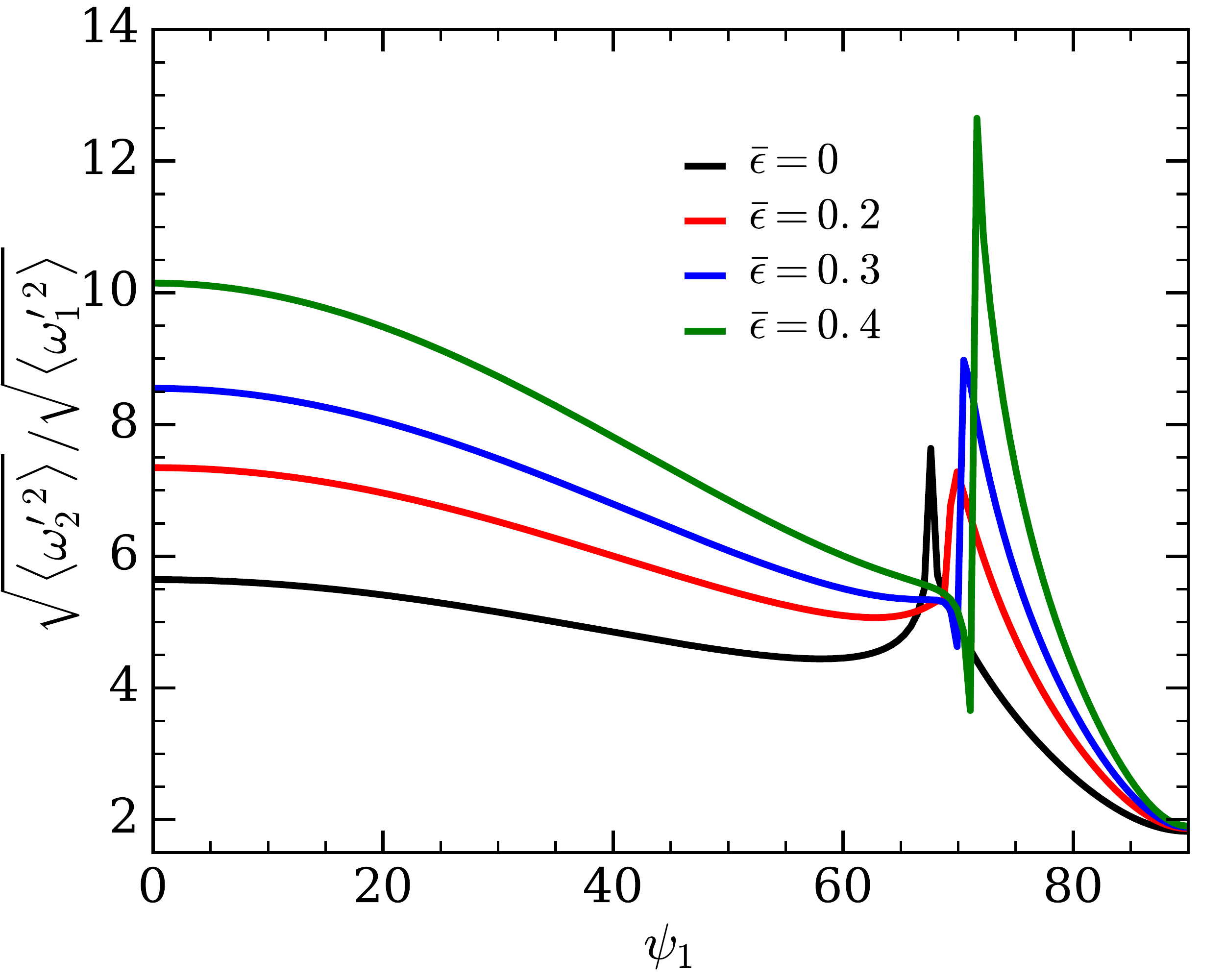}
  \caption{Amplification of vorticity across the shock as a function of angle $\psi$ for various values of nuclear dissociation parameter $\bar{\epsilon}$ for incident vorticity waves for ${\cal M}_1=5$. 
  \label{fig:w2w1}}
\end{figure}

In the limit of small incidence angle $\psi_1$, Eq.~(\ref{eq:omega2_scaling}) yields $\omega_2' \sim {\mathcal C} \omega_1'$. For a $\mathcal{M}_1=5$ shock, ${\mathcal C}\simeq5.6$, which is precisely what we observe in Fig.~\ref{fig:w2w1_phi} for $\sqrt{\langle\omega_2'^2\rangle}/\sqrt{\langle\omega_1'^2\rangle}$ for all the three curves. As predicted by Eq.~(\ref{eq:omega2_scaling}), the three curves gradually diverge with increasing $\psi_1$. Figure~\ref{fig:w2w1} shows the vorticity amplification $\sqrt{\langle\omega_2'^2\rangle}/\sqrt{\langle\omega_1'^2\rangle}$ for a incident vorticity wave across the shock as a function of angle $\psi_1$ for various values of nuclear dissociation parameter $\bar{\epsilon}$ for $\mathcal{M}_1=5$. Due to larger compression with increasing $\bar{\epsilon}$, the amplification $\sqrt{\langle\omega_2'^2\rangle}/\sqrt{\langle\omega_1'^2\rangle}$ also grows with $\bar{\epsilon}$, in agreement with the prediction of Eq.~(\ref{eq:omega2_scaling}). 

\begin{figure}
 \includegraphics[angle=0,width=1\columnwidth, clip=false]{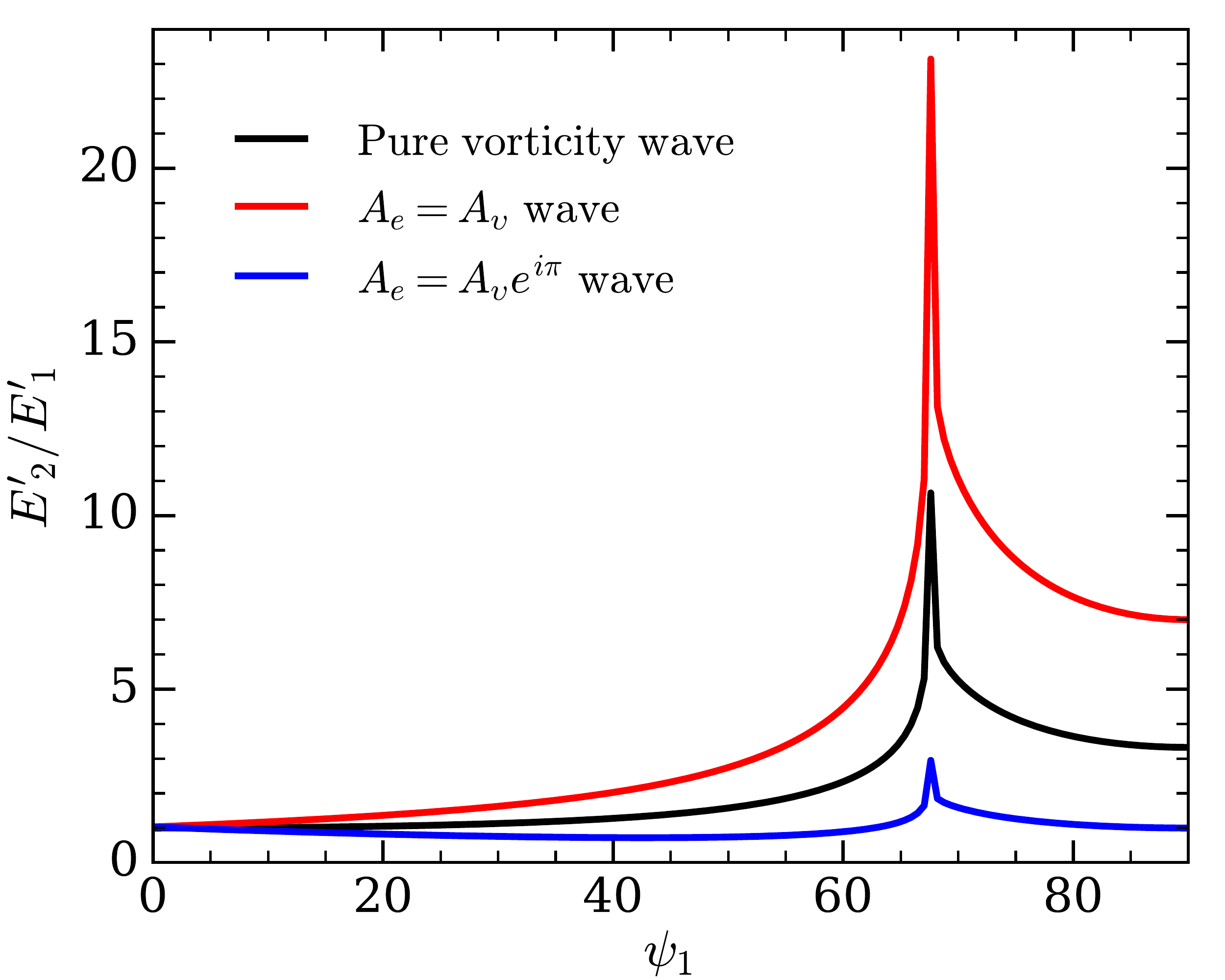}
  \caption{The amplification of turbulent kinetic energy across the shock with $\mathcal{M}_1=5$ as a function of angle $\psi$ for purely vorticity wave, in-phase vorticity-entropy wave (i.e., $A_e=A_\upsilon$), and out-of-phase vorticity-entropy wave (i.e., $A_e=A_\upsilon e^{i\pi}$). When the vorticity and entropy waves are in phase, we get the strongest amplification. When they are out of phase, there is no significant amplification.
  \label{fig:q2q1_phi}}
\end{figure}

\begin{figure}
 \includegraphics[angle=0,width=1\columnwidth, clip=false]{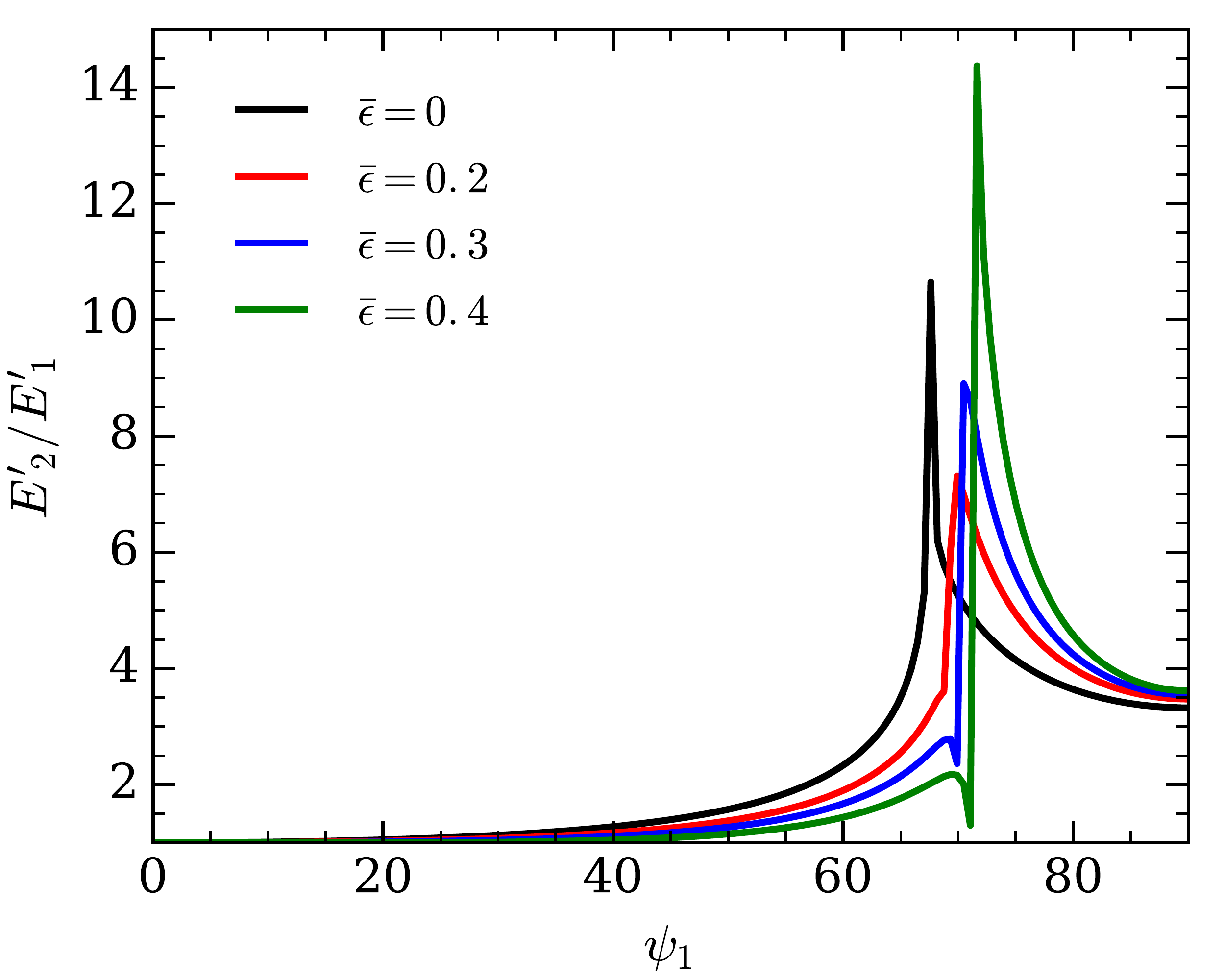}
  \caption{Amplification of turbulent kinetic energy across the shock as a function of angle $\psi_1$ for various values of nuclear dissociation parameter $\bar{\epsilon}$ for ${\cal M}_1=5$ for incident vorticity waves.
  \label{fig:q2q1}}
\end{figure}

\begin{figure}
 \includegraphics[angle=0,width=1\columnwidth, clip=false]{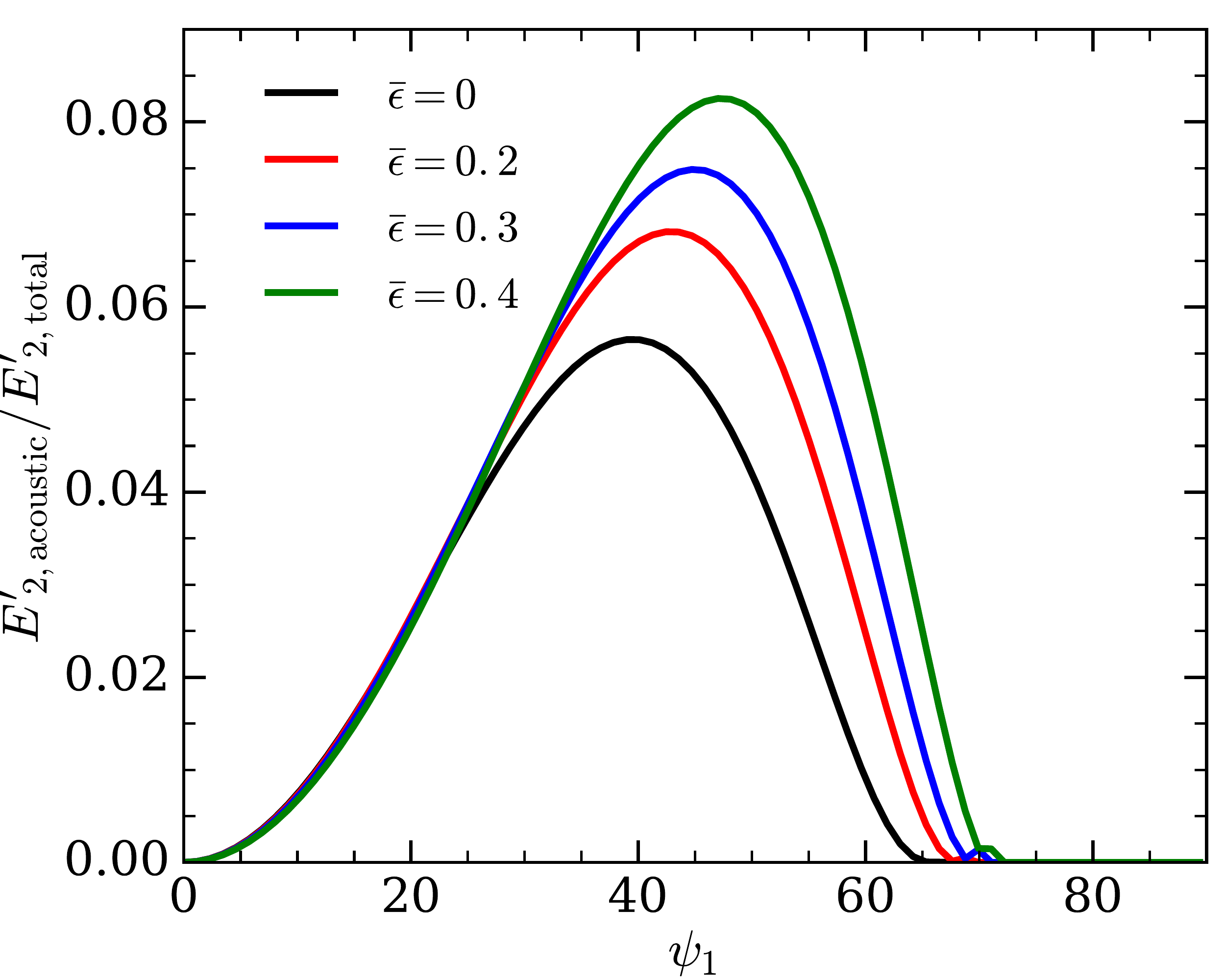}
  \caption{The ratio of kinetic energy association with sound wave to the total kinetic energy of the fluctuating field in the post-shock region as a function of the incidence angle $\psi_1$ for ${\cal M}_1=5$ for incident vorticity waves. 
  \label{fig:q2q1_ac_ratio}}
\end{figure}

Figure~\ref{fig:q2q1_phi} shows the ratio of turbulent kinetic energy $E'$ across the shock $E'_2/E'_1$ as a function of angle $\psi_1$ for $\mathcal{M}_1=5$ and $\bar\epsilon=0$ for the same three types of incident perturbations. We define $E'$ as 
\begin{equation}
E'=\frac{1}{2}\left(\langle u'u'^*\rangle+\langle \upsilon'\upsilon'^* \rangle \right),
\label{eq:eturb}
\end{equation}
where $\langle\rangle$ means averaging over $t$ and $y$, while sign ${}^*$ denotes complex conjugate. Similarly to the case of $\langle\omega'^2\rangle$, we observe the largest (smallest) $E'_2/E'_1$ when entropy and vorticity waves are in phase (out of phase), while for pure vorticity wave, $E'_2/E'_1$ is roughly the average of the two cases. For example, at $\psi_1=60^\circ$, for in-phase waves, we get $E'_2/E'_1=4.51$, while for out-of-phase waves, we get $E'_2/E'_1=0.90$, which means that the total kinetic energy of the perturbations actually decreases across the shock in this case for this value of $\psi_1$. For pure vorticity wave, we get $E'_2/E'_1=2.36$ for the same $\psi_1$. 

At $\psi_1\simeq\! 0$, we see no amplification of $E'$, while
the vorticity, as shown above, scales as
$\sqrt{\langle\omega_2'^2\rangle}/\sqrt{\langle\omega_1'^2\rangle} \sim
\mathcal{C}$. This is due to the fact that in this limit, the
$x$-component of the velocity perturbation $u_1'$ is $\simeq 0$, while
$\upsilon_1'\simeq A_\upsilon$, i.e., the velocity perturbation has only
$y$-component, which is tangential to the shock. The tangential
component of the velocity does not changes across the shock
\citep{landau:59}. Hence, there is no amplification of turbulent kinetic
energy in this limit. On the other hand, the vorticity $\omega'$ still
changes because it depends on the wavelength, which decreases by a
factor given by Eq.~(\ref{eq:kappa_ratio}).

The $x$-component of velocity of incident vorticity wave grows with
$\psi_1$ as $u_1' \propto \sin\psi_1$ (cf. Eq.~\ref{eq:u1}). The shock
responds sensitively to $u_1'$ by changing its position and shape
according to Eq.~(\ref{eq:shock_lia0}). Due to the deformation of the
shock, both $x$- and $y$-components of velocity will be perpendicular to
the shock at some $y$ and $t$. In this case, both $u'$ and $\upsilon'$ undergo
significant amplifications across the shock. The amplification factor
gradually grows with $\psi_1$ reaching, e.g., $\simeq\! 1.4$ and
$\simeq\! 2.3$ for $\psi_1=45^\circ$ for purely vorticity and in-phase
entropy-vorticity waves, respectively. The largest amplification is reached at $\psi_1 \simeq \psi_\mathrm{c}$, with amplification factors of $\simeq\! 6.2$ and $\simeq\! 13$ for these
two cases. However, such a large amplification is confined to a
narrow range of values of $\psi_1$ around $\psi_\mathrm{c}$ with width
$\lesssim 5^\circ$.

Figure~\ref{fig:q2q1} shows the amplification of kinetic energy as a function of $\psi_1$ for various values of nuclear dissociation parameter $\bar\epsilon$, ranging from $0$ to $0.4$. As we can see, $E'_2/E'_1$ exhibits only minor change with $\bar\epsilon$. The spike in $E'_2/E'_1$ around $\psi_1\simeq\!69^\circ$, which corresponds to the critical angle $\psi_1=\psi_c$, shifts towards slightly higher $\psi$ with $\bar\epsilon$ due to the fact the $\psi_c$ increases with $\bar\epsilon$ (cf. Fig.~\ref{fig:c}). 

Figure~\ref{fig:q2q1_ac_ratio} shows the ratio of the kinetic energy of the acoustic component to the total kinetic energy of the entire fluctuating velocity field as a function of the incidence angle $\psi_1$ for $\bar\epsilon=0,\ 0.2,\ 0.3,$ and $0.4$ for ${\cal M}_1=5$. This ratio can reach up to $0.08$ around $\psi\sim 50^\circ$, which is a non-negligible amount. 

\subsection{Interaction with Turbulence}
\label{sec:turbulence}

So far, our analysis has focused on interaction of shocks with
individual fluctuation modes. In the following, we consider interaction
with turbulent fields, which we model as sets of
random three-dimensional vorticity and entropy
waves. In the LIA, each of these waves interact independently
with the shock. The full turbulent statistics behind the shock can be
obtained by integrating over the interactions of each of these waves
with the shock.

In order to achieve this goal, we first need to establish how the 2D LIA presented so far is
related to the general three-dimensional problem. In 3D Cartesian
coordinate system $(x,y,z)$, consider an incident planar wave with
wavenumber $\vec{\kappa}_1$ that makes angle $\psi_1$ with the $x$-axis.
The latter is assumed to be perpendicular to the shock. The dynamics in
the plane spanned by vector $\vec{\kappa}_1$ and the shock normal is
identical to that of the 2D LIA problem. The component of the velocity
field perpendicular to this plane passes unchanged through the shock,
while the components parallel to the plane change according to LIA
\citep{ribner:54,mahesh:96,wouchuk:09}. In the following, we refer to
this plane as the LIA plane.

We consider two types of turbulent fields. The first is an anisotropic turbulence characterized by relation
\begin{equation}
R_{rr}=R_{\theta\theta}+R_{\phi\phi} \quad \mathrm{and} \quad R_{\theta\theta}=R_{\phi\phi},
\label{eq:rrr}
\end{equation}
where $R_{ij}$ is the $ij$-component of the Reynolds stress tensor. This means an equipartition between radial and non-radial components of turbulent kinetic energy and it was observed in buoyancy-driven turbulent convection in stellar interiors~\citep[e.g.,][]{arnett:09}. The second type is a fully isotropic turbulence represented by
\begin{equation}
R_{rr}=R_{\theta\theta}=R_{\phi\phi}.
\label{eq:rrr2}
\end{equation} 
Estimate of how well Eq.~(\ref{eq:rrr2}) describes the turbulence in stellar convective shells is beyond the scope of this work. Instead, we use it as an alternative prescription in order to test the sensitivity of our results to the properties of upstream turbulence.

In order to model turbulence characterized by these relations, we randomly sample the velocity field $(v_x, v_y, v_z)$ with a statistics that satisfies these relations. Here, the $x$-component $v_x$ plays a role similar to that of the radial component in CCSNe since both of them are perpendicular to the shocks in their respective contexts. By the same rationale, $v_y$ and $v_z$ play the roles of angular components in CCSNe. 

The wavenumber vectors $\vec{\kappa}_1$ of incident waves are sampled
randomly with uniform distribution on a 2D sphere. For each wave, we
decompose the velocity field into three components: the first being
perpendicular to the LIA plane, the second being perpendicular to
$\vec{\kappa}_1$ on the LIA plane, and the third being parallel to
$\vec{\kappa}_1$ on the LIA plane. The first component passes through
the shock unchanged, while the second changes according to LIA. The
third component represents the non-solenoidal part of the
velocity field, which we set to zero when constructing the
vorticity waves.

Using this velocity field, we first investigate how the spectrum of turbulence changes as it crosses the shock. Each turbulent eddy is characterized by its wavenumber $\vec{\kappa}$. According to LIA, when a turbulent eddy with a wavenumber $\vec{\kappa}_1$ passes through the shock, the $x$-component of $\vec{\kappa}_1$ increases from $\kappa_{1,x}$ to ${\cal C} \kappa_{1,x}$. The other two components, $\kappa_{1,y}$ and $\kappa_{1,z}$, do not change. Thus, the wavenumber vector increases from 
\begin{equation}
\kappa_1 = \sqrt{\kappa_{1,x}^2+\kappa_{1,y}^2+\kappa_{1,z}^2}
\end{equation}
to
\begin{equation}
\kappa_2 = \sqrt{{\cal C}\kappa_{1,x}^2+\kappa_{1,y}^2+\kappa_{1,z}^2}=\sqrt{({\cal C}-1)\kappa_{1,x}^2+\kappa_1^2}. 
\end{equation}
Hence,
\begin{equation}
\label{eq:kratio}
\frac{\kappa_2}{\kappa_1} =\sqrt{({\cal C}-1)\cos^2\psi_1+1},
\end{equation}
where we used the definition $\cos\psi_1=\kappa_{1,x}/\kappa_1$. Since $\kappa = 2\pi/\lambda$, where $\lambda$ is the spatial scale of our eddy, the eddy becomes smaller by factor $\kappa_2/\kappa_1$ as it passes through the shock. At most, $\lambda$ can decrease by a factor of ${\cal C}$, which happens when $\vec{\kappa}_1$ is perpendicular to the shock. When $\vec{\kappa}_1$ is parallel to the shock, there is no change in the size of the eddy. 

\begin{figure}
 \includegraphics[angle=0,width=1\columnwidth, clip=false]{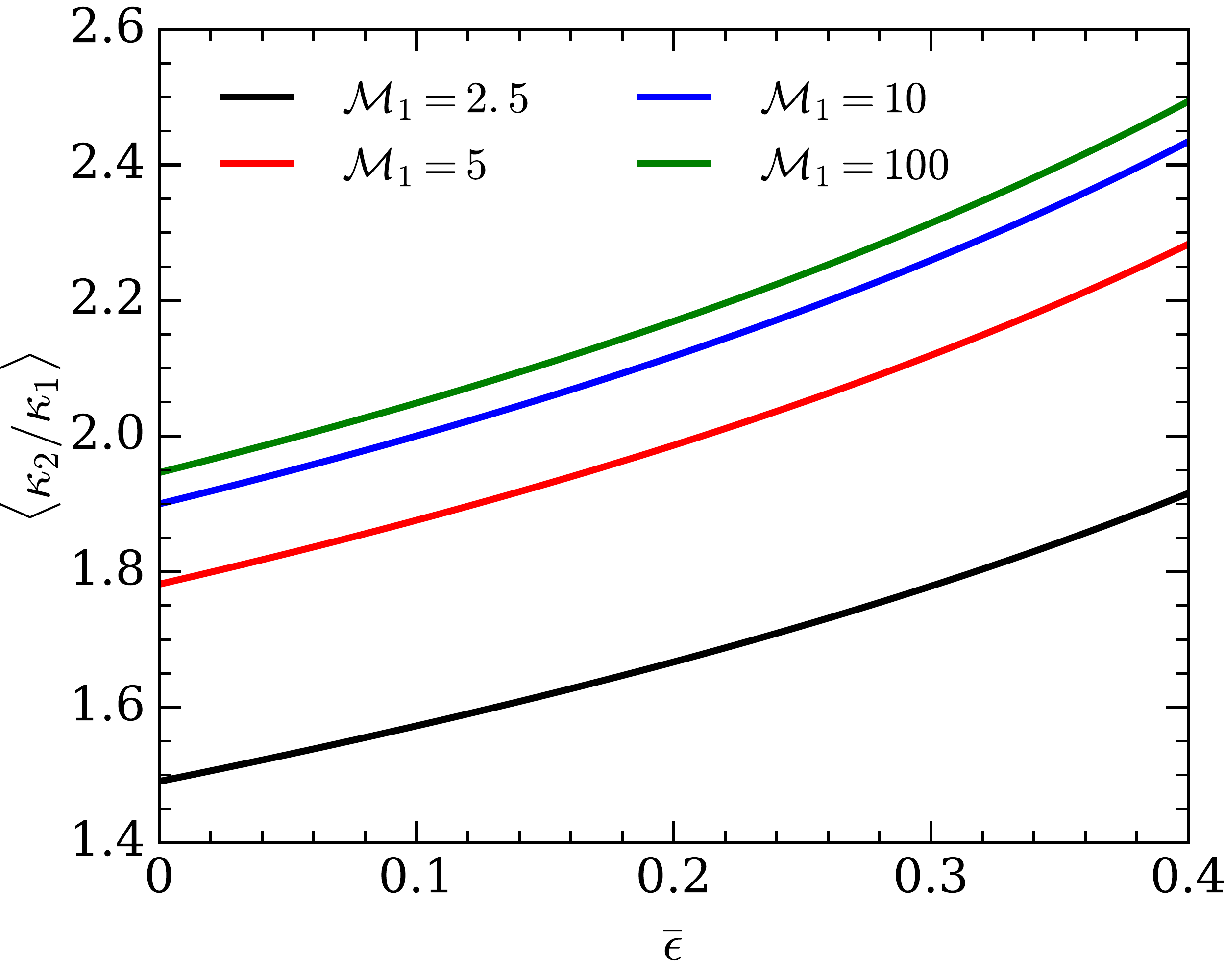}
  \caption{The average ratio of downstream and upstream wavenumbers of incident vorticity and/or entropy waves as a function of nuclear dissociation parameter $\bar\epsilon$ for various values of upstream Mach number ${\cal M}_1$. For all the values of $\bar\epsilon$ and ${\cal M}_1$ considered here, the average wavenumber of the upstream turbulent field increases as it crosses the shock, meaning that the spectrum of the turbulent motion shifts towards smaller wavelengths.
  \label{fig:kratio}}
\end{figure}

\begin{figure}
 \includegraphics[angle=0,width=1\columnwidth, clip=false]{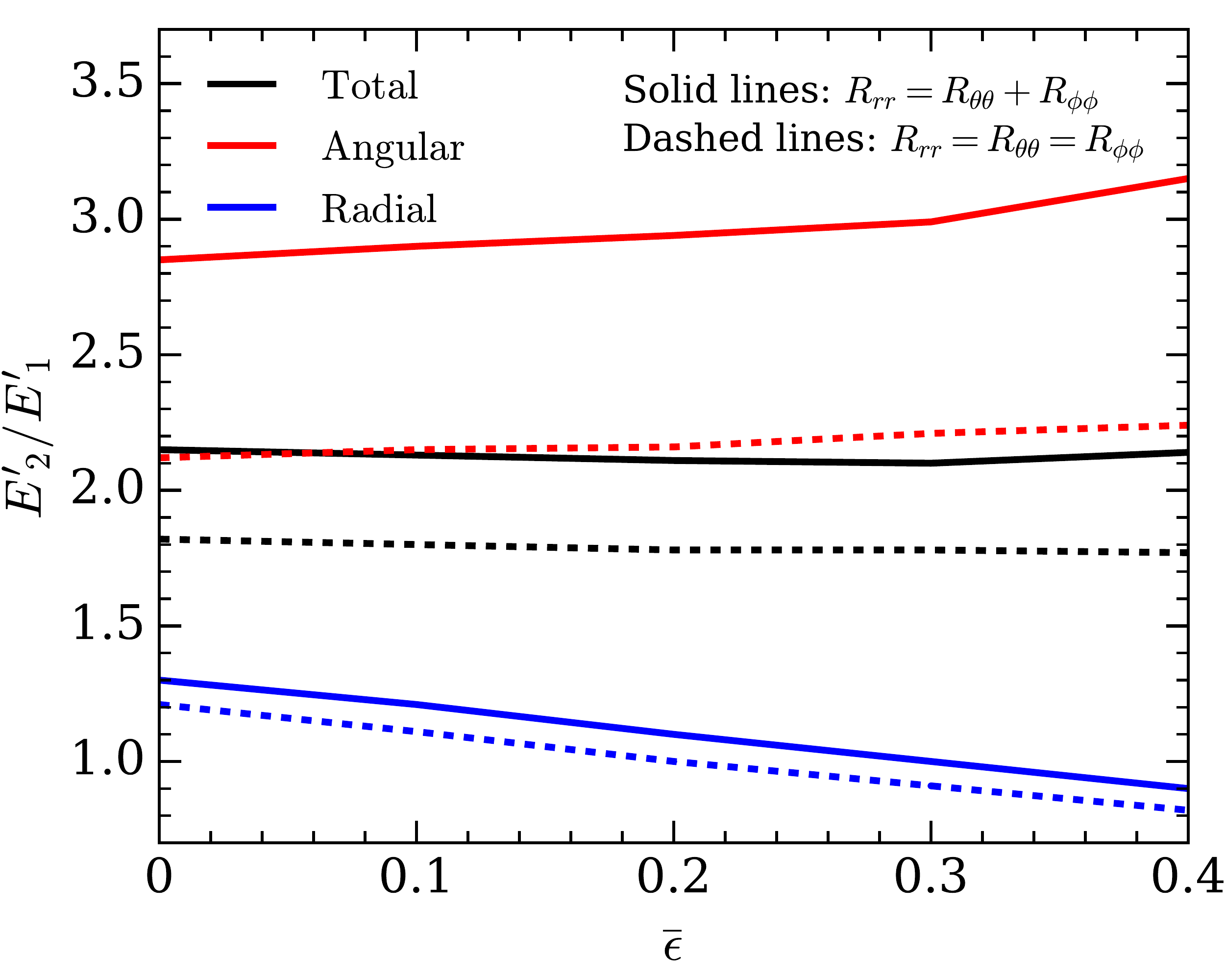}
  \caption{The amplification of turbulent kinetic energy across the shock as a function of nuclear dissociation parameter $\bar\epsilon$ for incident vorticity waves. The black line represents the amplification of the total kinetic energy, while red and blue lines represent the amplifications of angular and radial components of the kinetic energy. The solid lines correspond to anisotropic turbulence represented by relation $R_{rr}=R_{\theta\theta}+R_{\phi\phi}$, while the dashed lines correspond to fully isotropic turbulence represented by relation $R_{rr}=R_{\theta\theta}=R_{\phi\phi}$. \label{fig:eratioA0}}
\end{figure}

\begin{figure}
 \includegraphics[angle=0,width=1\columnwidth, clip=false]{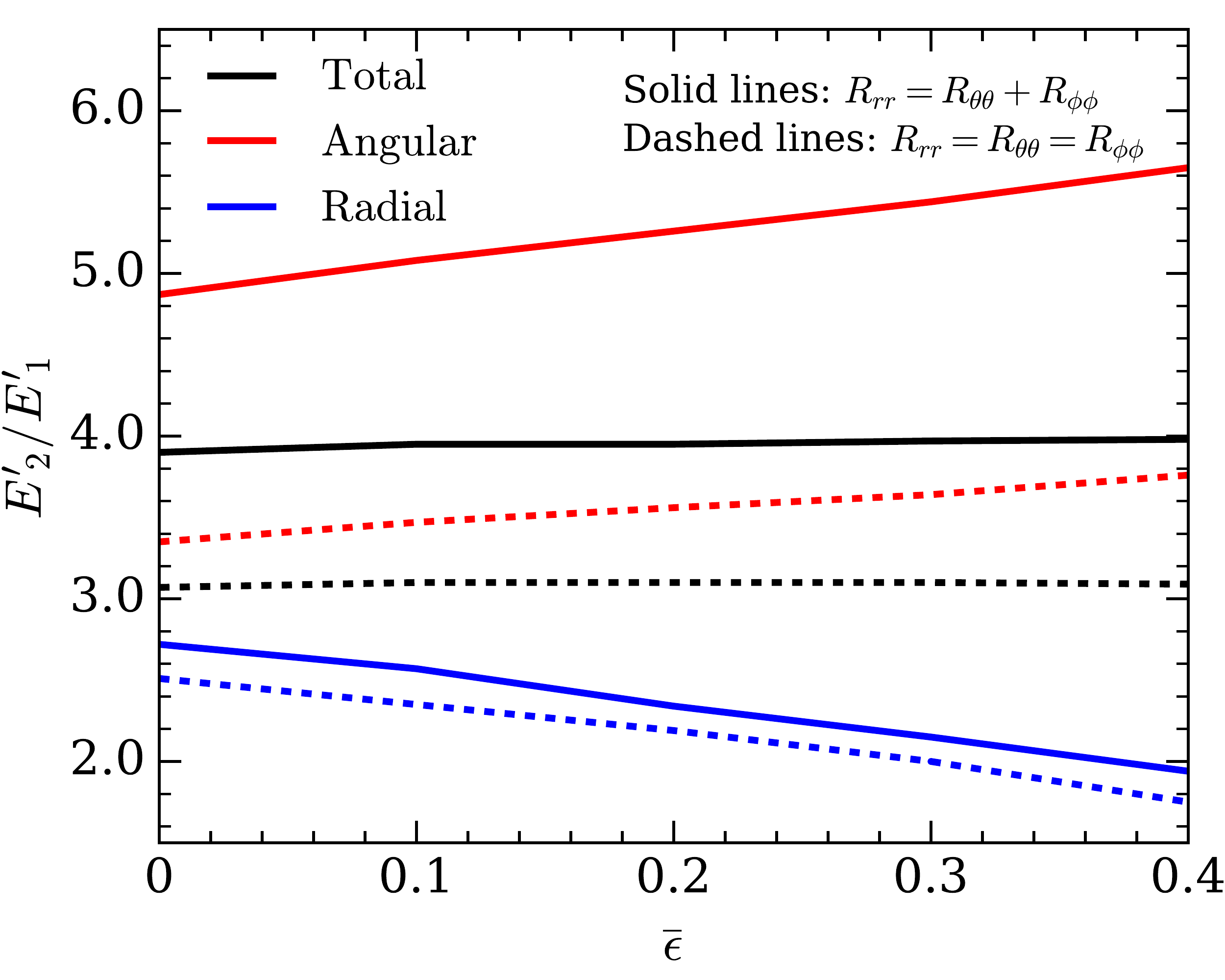}
  \caption{The amplification of turbulent kinetic energy across the shock as a function of nuclear dissociation parameter $\bar\epsilon$ for incident entropy vorticity waves of the same phase. The black line represents the amplification of the total kinetic energy, while red and blue lines represent the amplifications of angular and radial components of the kinetic energy. The solid lines correspond to anisotropic turbulence represented by relation $R_{rr}=R_{\theta\theta}+R_{\phi\phi}$, while the dashed lines correspond to fully isotropic turbulence represented by relation $R_{rr}=R_{\theta\theta}=R_{\phi\phi}$. \label{fig:eratioA1}}
\end{figure}

\begin{figure}
 \includegraphics[angle=0,width=1\columnwidth, clip=false]{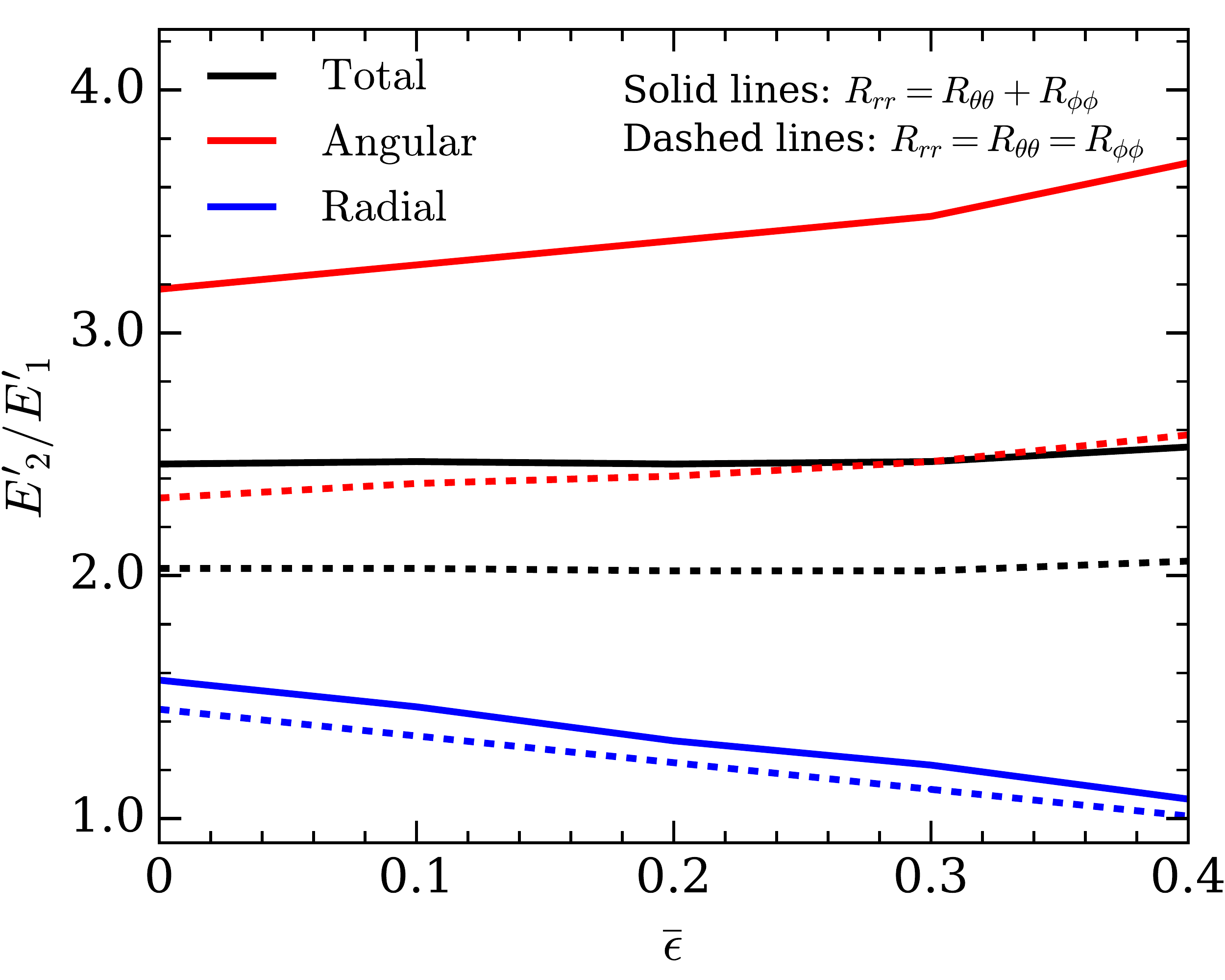}
  \caption{The amplification of turbulent kinetic energy across the shock as a function of nuclear dissociation parameter $\bar\epsilon$ for incident entropy and vorticity waves. The phase difference between the two waves are chosen randomly with a uniform distribution in $(0,2\pi)$. The black line represents the amplification of the total kinetic energy, while red and blue lines represent the amplifications of angular and radial components of the kinetic energy. The solid lines correspond to anisotropic turbulence represented by relation $R_{rr}=R_{\theta\theta}+R_{\phi\phi}$, while the dashed lines correspond to fully isotropic turbulence represented by relation $R_{rr}=R_{\theta\theta}=R_{\phi\phi}$. \label{fig:eratio}}
\end{figure}

In order to obtain an average behavior of $\kappa_2/\kappa_1$, we average it over a random set of vectors $\vec{\kappa}_1$ with uniform distribution on a 2D sphere. This is equivalent to sampling $\cos\psi_1$ uniformly in interval $[0,1]$, which, in turn, is equivalent to solving integral
\begin{equation}
\langle \frac{\kappa_2}{\kappa_1} \rangle = \int_0^1 \sqrt{({\cal C}-1)x+1 } \ dx.
\end{equation}
The latter can be calculated analytically:
\begin{equation}
\langle \frac{\kappa_2}{\kappa_1} \rangle = \frac{2}{3}\frac{1+\sqrt{\cal C}+{\cal C}}{1+\sqrt{\cal C}},
\end{equation}
Figure~\ref{fig:kratio} shows the average ratio $\langle \kappa_2/\kappa_1 \rangle$ as a function of nuclear dissociation parameter $\bar\epsilon$ for four value of upstream Mach number: 2.4, 5, 10, and 100. In all cases, $\langle \kappa_2/\kappa_1 \rangle$ increases mildly with $\bar \epsilon$. This is a simple reflection of the fact that the compression factor increases with $\bar \epsilon$, as we discussed above. For our fiducial values of $\bar\epsilon=0.2$ and ${\cal M}_1=5$, $\langle \kappa_2/\kappa_1 \rangle \simeq 2$. As expected, this result does not change much with further increasing ${\cal M}_1$. 

We note that this decrease in the radial extent may be partially compensated and possibly offset by the previous radial stretching experienced by the turbulent fluctuations during the collapse \citep{lai:00n,takahashi:14}. 

Now consider the post-shock turbulent kinetic energy. The total specific kinetic energy for a single wave is given by Eq.~(\ref{eq:eturb}), which we rewrite as
\begin{equation}
E' = \frac{1}{2}\left(\langle u'^2 \rangle + \langle \upsilon'^2 \rangle \right),
\end{equation}
where $\langle u'^2 \rangle = \langle u' u'^* \rangle$. For an incident vorticity-entropy wave of form (\ref{eq:T1}), we obtain
\begin{equation}
E'_1 = \frac{1}{2}U_1^2 |A_\upsilon|^2,
\end{equation}
For the downstream vorticity field (\ref{eq:u2})-(\ref{eq:v2}), in the far-field region ($x \gg1/\kappa$), we have\footnote{Note that Eq.~(\ref{eq:eturb2}) does not include the contribution from acoustic waves. In order to include that, have to add $\frac{1}{2}U_1^2 \left(|F|^2+|H|^2 \right)$ to the right-hand side of Eq.~(\ref{eq:eturb2}) in the propagative regime ($\psi_1<\psi_\mathrm{c}$). In the non--propagative regime, there is no contribution from the acoustic component.}
\begin{equation}
\label{eq:eturb2}
E'_2 =\frac{1}{2} U_1^2 \left(|G|^2+|I|^2 \right),
\end{equation}
Thus, the ratio of upstream and downstream turbulent kinetic energies are
\begin{equation}
\frac{E'_2}{E'_1} = |\tilde G|^2+|\tilde I|^2 , 
\label{eq:eturbratio}
\end{equation}
where $\tilde G=G/A_\upsilon$ and  $\tilde I=I/A_\upsilon$. 

Note that formula (\ref{eq:eturbratio}) depends only on $A_e/A_\upsilon$ and the incidence angle $\psi_1$ of upstream vorticity-entropy waves, but not on their wavenumbers $\kappa_1$. This is an important result because it means that the amplification factor of turbulent kinetic energy across the shock is independent of the spectrum of upstream turbulence.

The black line in Fig.~\ref{fig:eratioA0} shows the amplification of the total kinetic energy across the shock as a function of the nuclear dissociation parameter $\bar \epsilon$ for incident vorticity waves. Here, we use an anisotropic turbulent field represented by relation (\ref{eq:rrr}) and each point on this graph is calculated using a sample of 150,000 random incident waves. The amplification of the total energy $E'_2/E'_1$ does not change much with $\bar\epsilon$, remaining at $\simeq\! 2.14$ as $\bar\epsilon$ grows from $0$ to $0.4$. On the other hand, the amplification of the angular and radial components, defined as $(E'_{y,2}+E'_{z,2})/(E'_{y,1}+E'_{z,1})$ and $E'_{x,2}/E'_{x,1}$, exhibit noticeable dependence on $\bar\epsilon$. The angular component, shown with the red line in Fig.~\ref{fig:eratioA0}, increases from $2.85$ to $3.15$ as $\bar\epsilon$ grows from $0$ to $0.4$. Contrary to this, the amplification of the radial component, shown with the blue line in Fig.~\ref{fig:eratioA0}, decreases from $1.30$ to $0.90$ for the same values of $\bar\epsilon$.

Similar to the behavior of individual waves discussed earlier in Section~\ref{sec:wave}, the change of kinetic energy of the incident vorticity waves across the shock is very sensitive to the presence of incident entropy waves. If we add entropy waves with the same phase and amplitude as the incident vorticity waves (i.e., $A_e=A_\upsilon$), the amplification of total kinetic energy of turbulent field becomes $\simeq\! 3.95$ (cf. the dashed black line in Fig.~\ref{fig:eratioA1}). This is $\simeq\! 1.85$ times larger than what we get in the case of pure vorticity waves shown in Fig.~\ref{fig:eratioA0}. On the other hand, if they are out of phase (i.e., $A_e=A_\upsilon e^{i\pi}$), we find that the total energy does not change much and $E'_2/E'_1 \sim 1$ (not shown here). Such a dependence on entropy waves is a direct manifestation of the simple scaling law (\ref{eq:omega2_scaling}) discussed above.

We also consider the case when the phase difference between the incident entropy and vorticity waves are chosen randomly with uniform distribution between $0$ and $2\pi$. This case is presented in Fig.~\ref{fig:eratio}. We find that in this case, the overall behavior of the turbulent kinetic energy is qualitatively and quantitatively similar to that in the case of incident pure vorticity waves shown in Fig.~\ref{fig:eratioA0}. For fiducial parameters, ${\cal M}_1=5$ and $\bar\epsilon=0.2$, we get $E'_2/E'_1=2.46$, $(E'_{y,2}+E'_{z,2})/(E'_{y,1}+E'_{z,1}) = 3.38$, and $E'_{x,2}/E'_{x,1}=1.32$.

We can summarize these findings as follows. If the phases of incident entropy and vorticity are strongly correlated, then the total kinetic energy of the turbulent field will increase by a factor of $\sim\! 4$. If they are strongly anti-correlated, then there is no amplification. If there is no correlation in the phases, then the amplification is $\sim\! 2$.

In order to test the sensitivity of our results to the particular form of (\ref{eq:rrr}), we repeat this exercise for isotropic turbulence represented by Eq.~(\ref{eq:rrr2}). The dashed black lines in Figs.~\ref{fig:eratioA0} and \ref{fig:eratioA1} show the amplification of the total turbulent kinetic energy of the field of incident vorticity and in-phase entropy-vorticity waves, respectively. In both cases, the amplification is again insensitive to $\bar\epsilon$, remaining at $\simeq\! 1.8$ and $\simeq\! 3.1$ for incident vorticity and in-phase entropy-vorticity waves. These values are $\simeq\! 15\%$ and $\simeq\! 21\%$ smaller than those in the case of anisotropic turbulence. Despite similarity of the of the behavior of the total energy across the shock, we see differences in the behavior of radial and angular components.

For isotropic turbulence, the amplification factors of the angular and radial components are closer to each other than those for the anisotropic turbulence. For example, for $\bar\epsilon=0.2$ and incident vorticity wave, the ratios of the amplifications of angular and radial components are $2.67$ and $2.16$ for anisotropic and isotropic turbulence models. The reason for larger amplification of angular component in anisotropic turbulence is due fact that a smaller fraction of the kinetic energy is contained in a component tangential to the shock, which does not undergo amplification. 

\begin{figure}
 \includegraphics[angle=0,width=1\columnwidth, clip=false]{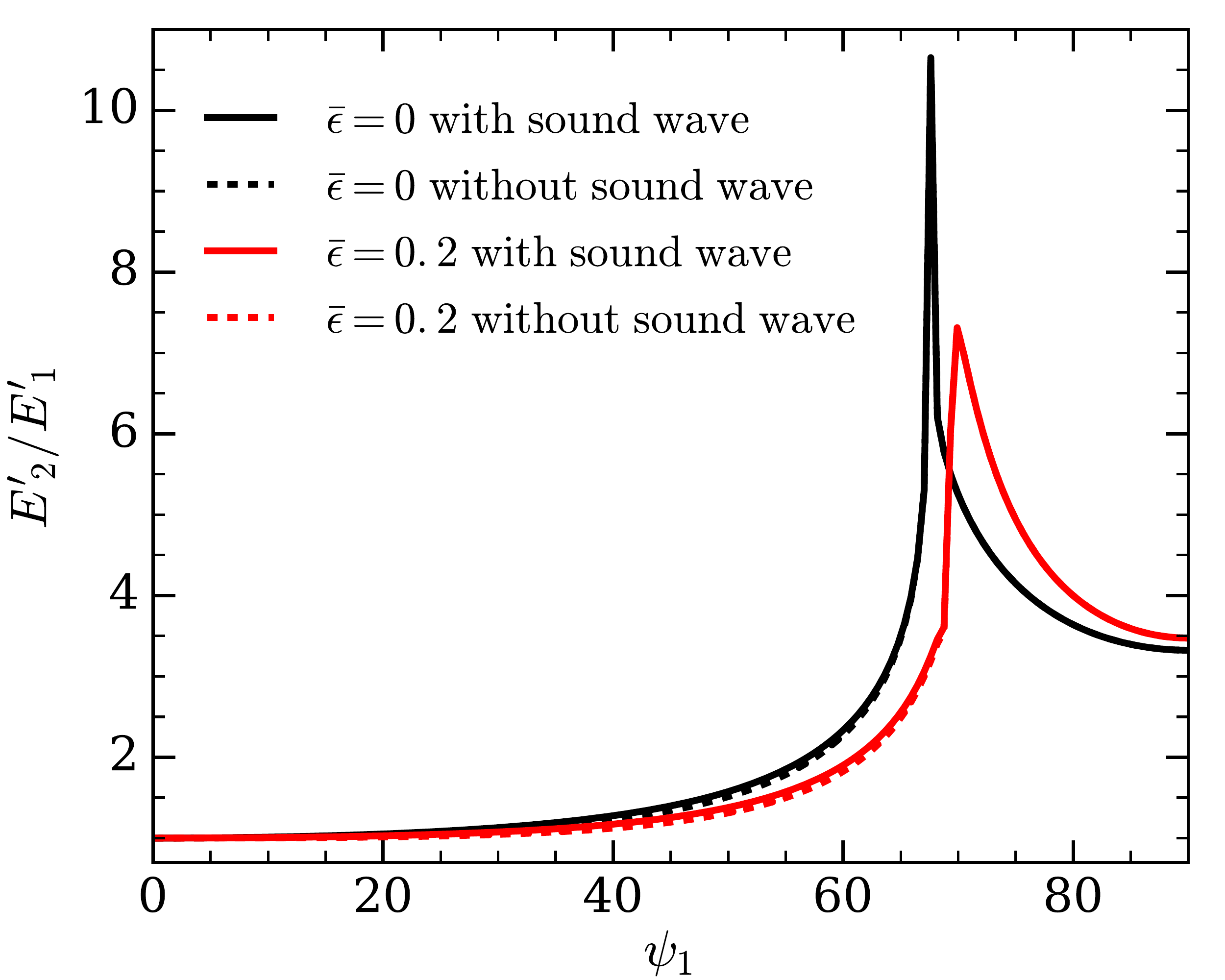}
  \caption{The ratios of upstream and downstream kinetic energies of turbulent field as a function of the incidence angle $\psi_1$ for ${\cal M}_1=5$ for incident vorticity waves. The solid lines include the contribution of sound waves to the kinetic energy, while the dashed lines do not. 
  \label{fig:q2q1_ac_phi}}
\end{figure}

Our analysis shows that, downstream of the shock, acoustic waves contribute at most $\sim 2\%$ of the total turbulent kinetic energy, which is a tiny amount. This may seem surprising in the
light of the fact that the ratio of the kinetic energy of sound waves to
the kinetic energy of the total fluctuating velocity field can reach
$\sim 0.08$ for $\psi \sim 50^\circ$, as we saw in
Fig.~\ref{fig:q2q1_ac_ratio}. However, the total kinetic energy of the
fluctuating field in this region is small compared to that at larger
$\psi$. This is easily visible in Fig~\ref{fig:q2q1_ac_phi}, which shows, for incident vorticity waves,
the ratio of upstream and downstream kinetic energies with and without
the contribution of the downstream acoustic field with solid and dashed
lines, respectively. As we can see, if we average over all values of
$\psi_1$, the contribution of the acoustic component should be
negligibly small, in agreement with our findings above. 

\begin{figure}
 \includegraphics[angle=0,width=1\columnwidth, clip=false]{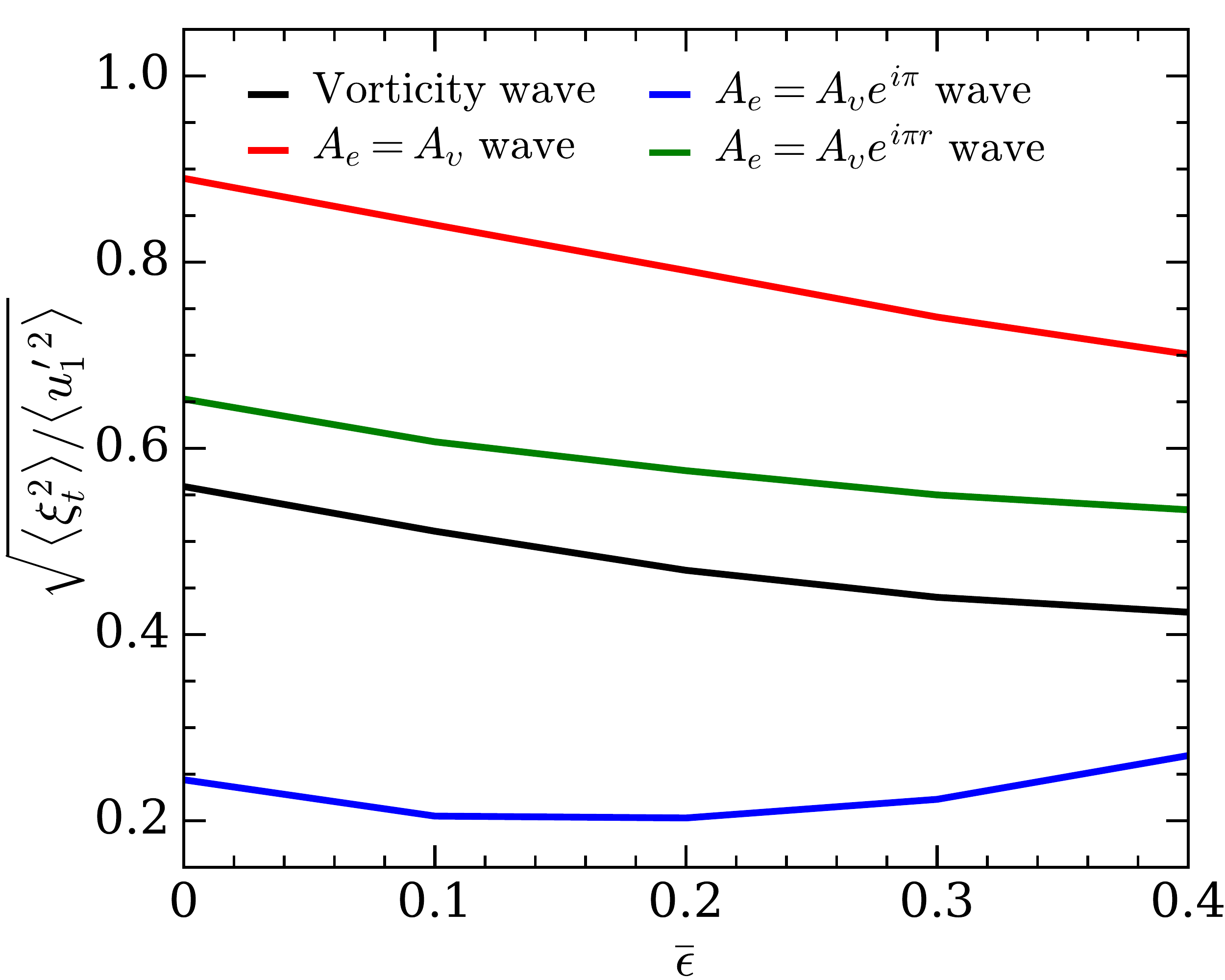}
  \caption{The normalized RMS velocity of the shock oscillations $\sqrt{\langle \xi_t^2 \rangle / \langle u_1'^2 \rangle}$ as a function of nuclear dissociation parameter $\bar\epsilon$ for ${\cal M}_1=5$. The black line corresponds to incident vorticity waves, while the red and blue lines correspond to incident entropy and vorticity waves with the same phase (i.e., $A_e=A_\upsilon$) and $180^\circ$ phase difference ($A_e=A_\upsilon e^{i\pi}$), respectively. Finally, the green line represents the case where the entropy and vorticity waves have randomly sampled phase differences from $0$ to $2\pi$ with uniform distribution (i.e., $A_e=A_\upsilon e^{i\pi r}$ where $r \in [0,2]$ is random number). We get the strongest velocities $\sqrt{\langle \xi_t^2 \rangle / \langle u_1'^2 \rangle}$ of $\sim 0.8$ in the case of in-phase entropy-vorticity waves. When they are out of phase, we get the weakest amplification $\sqrt{\langle \xi_t^2 \rangle / \langle u_1'^2 \rangle} \sim 0.2$. In the case of randomly distributed phase differences and in the case of incident vorticity waves, we get similar values of $\sqrt{\langle \xi_t^2 \rangle / \langle u_1'^2 \rangle}$ of $\sim 0.58$ and $\sim 0.47$, respectively. 
  \label{fig:xit}}
\end{figure}

The shock surface responds to upstream velocity perturbations by oscillating according to formula (\ref{eq:shock_lia0}). Fig.~\ref{fig:xit} shows the normalized RMS velocity of the shock $\sqrt{\langle \xi_t^2 \rangle / \langle u_1'^2 \rangle}$ as a function of nuclear dissociation parameter $\bar\epsilon$ for ${\cal M}_1=5$. Here, $\langle u_1'^2 \rangle$ is the RMS value of the $x$-component of the perturbation velocity. The black line corresponds to incident vorticity waves, while the red and blue lines correspond to incident entropy and vorticity waves with the same phase (i.e., $A_e=A_\upsilon$) and $180^\circ$ phase difference ($A_e=A_\upsilon e^{i\pi}$), respectively. Finally, the green line represents the case where the entropy and vorticity waves have randomly sampled phase differences from $0$ to $2\pi$ with uniform distribution (i.e., $A_e=A_\upsilon e^{i\pi r}$ where $r$ is a random number with uniform distribution in $[0,2]$). Similar to the amplification of turbulent kinetic energy, the shock velocity does not change much with $\bar\epsilon$, but it is very sensitive to the presence of entropy waves. For our fiducial value $\bar\epsilon=0.2$, we get the largest $\sqrt{\langle \xi_t^2 \rangle / \langle u_1'^2 \rangle}$ of $\sim 0.8$ for $A_e=A_\upsilon$, while for case $A_e=A_\upsilon e^{i\pi}$, we get the smallest $\sqrt{\langle \xi_t^2 \rangle / \langle u_1'^2 \rangle}$ of $\sim 0.2$. In the case of incident entropy-vorticity waves with randomly distributed phase differences and in the case of incident vorticity waves, we get similar values of $\sqrt{\langle \xi_t^2 \rangle / \langle u_1'^2 \rangle}$ $\sim 0.58$ and $\sim 0.47$, respectively. Note that, due to the employed normalization, these values do not depend on weather we use anisotropic (Eq. \ref{eq:rrr}) or isotropic (Eq. \ref{eq:rrr2}) turbulence prescriptions.

\section{Implications for the Explosion Condition}
\label{sec:implications}

We now discuss the implications of the above results on the conditions for producing explosion using the concept of critical luminosity \citep{burrows:93}. According to \citet{mueller:15}, in the presence of post-shock turbulence, the critical luminosity for producing explosion is
\begin{equation}
\label{eq:critlum}
L_\nu E_\nu^2 \propto (\dot M M)^{3/5}r_\mathrm{gain}^{-2/5}\left(1+\frac{4\langle {\cal M'}_2^2\rangle}{3}\right)^{-3/5},
\end{equation}
where $\langle {\cal M'}_2^2 \rangle$ is the RMS post-shock turbulent Mach number. Following \citet{mueller:15}, we define it as
\begin{equation}
\langle {\cal M'}^2 \rangle = \frac{\langle \upsilon_\mathrm{a}^2 \rangle}{\langle c_\mathrm{s}^2 \rangle},  
\end{equation}
where $\langle \upsilon_\mathrm{a}^2 \rangle$ is RMS angular velocity, which can be expressed in terms of specific kinetic energy of angular turbulent motion as $\langle \upsilon_\mathrm{a}^2 \rangle = 2E'_\mathrm{a}$. Using this, we can write 
\begin{equation}
\langle {\cal M'}_2^2 \rangle = \frac{2E'_\mathrm{a,2}}{\langle c_\mathrm{s}^2 \rangle} = 2 \frac{E'_\mathrm{a,2}}{E'_\mathrm{a,1}}\frac{E'_\mathrm{a,1}}{\langle c_\mathrm{s}^2 \rangle} = \frac{E'_\mathrm{a,2}}{E'_\mathrm{a,1}}\frac{\langle \upsilon_\mathrm{a,1}^2 \rangle}{\langle c_\mathrm{s,2}^2 \rangle}  = \frac{E'_\mathrm{a,2}}{E'_\mathrm{a,1}} \frac{\langle c_\mathrm{s,1}^2 \rangle}{\langle c_\mathrm{s,2}^2 \rangle} \langle {\cal M'}_1^2 \rangle
\end{equation}
Substituting this into Eq.~(\ref{eq:critlum}), we get
\begin{equation}
L_\nu E_\nu^2 \propto  \left(1+\frac{4}{3} 
\frac{E'_\mathrm{a,2}}{E'_\mathrm{a,1}} \frac{\langle c_\mathrm{s,1}^2 \rangle}{\langle c_\mathrm{s,2}^2 \rangle} \langle {\cal M'}_1^2 \rangle \right)^{-3/5} \simeq 1-\frac{4}{5} 
\frac{E'_\mathrm{a,2}}{E'_\mathrm{a,1}} \frac{\langle c_\mathrm{s,1}^2 \rangle}{\langle c_\mathrm{s,2}^2 \rangle} \langle {\cal M'}_1^2 \rangle,
\end{equation}
Note that we linearized in $\langle {\cal M'}_1^2 \rangle$ in the last step. Subtracting this from the critical luminosity in the absence of post-shock turbulence, we obtain an expression for the relative reduction of the critical luminosity due to upstream turbulence:
\begin{equation}
\label{eq:critlum1}
\delta (L_\nu E_\nu^2) \simeq \frac{4}{5} 
\frac{E'_\mathrm{a,2}}{E'_\mathrm{a,1}} \frac{\langle c_\mathrm{s,1}^2 \rangle}{\langle c_\mathrm{s,2}^2 \rangle} \langle {\cal M'}_1^2 \rangle,
\end{equation}
For our fiducial parameters $\bar\epsilon=0.2$ and ${\cal M}_1=5$, $\langle c_\mathrm{s,1}^2 \rangle / \langle c_\mathrm{s,2}^2 \rangle \simeq 0.25$ and $E'_\mathrm{a,2}/E'_\mathrm{a,1} \sim 3$ for anisotropic turbulence represented by relation (\ref{eq:rrr}) for an incident field of vorticity or entropy-vorticity waves with uncorrelated phases. For these values, Eq.~(\ref{eq:critlum1}) reduces to 
\begin{equation}
\label{eq:critlum2}
\delta (L_\nu E_\nu^2) \simeq 0.6 \langle {\cal M'}_1^2 \rangle,
\end{equation}
Thus, the critical luminosity decreases by $\simeq\!0.6 \langle {\cal M'}_1^2 \rangle$ compared to the case with no post-shock turbulence. 

In convective shells, we expect $\sqrt{\langle {\cal M'}^2 \rangle} \sim\!0.1$ \citep[e.g.,][]{mueller:16b}. During collapse, the Mach number of non-radial fluctuations increases as $\propto r^{(3\gamma-7)/4}$ \citep{lai:00n}. Assuming that convective shells fall from a radius of $\sim\!1500\,\mathrm{km}$ to $\sim\!200\,\mathrm{km}$ before it hits the shock, in the absence of turbulent dissipation, the turbulent Mach number should increase to $\sim 0.45$ before hitting the shock, which yields $\langle {\cal M'}_1^2\rangle\sim\!0.21$. This results in a reduction of the critical luminosity by $\sim\!12\%$ compared to the case with no upstream turbulence. 

Note that the estimate (\ref{eq:critlum2}) is of limited accuracy for a number of reasons. First, it neglects turbulent dissipation in the post-shock region. Second, it is based on a comparison to the hypothetical case with no post-shock turbulence. However, by the time a nuclear-burning shell hits the shock, the post-shock region is expected to have a fully-developed neutrino-driven turbulent convection \citep{couch:15b}, which we cannot include in our estimate. Both of these effects overestimate the reduction of critical luminosity. Therefore, the above estimate is expected to yield an approximate upper limit for the reduction of the critical luminosity.

\section{Conclusion}
\label{sec:summary}

In this paper, we studied the interaction of the shock waves in
core-collapse supernovae (CCSNe) with turbulent convection arising from
nuclear shell burning. We used a first-order perturbation theory called
the linear interaction approximation (LIA), which we extended to include
nuclear dissociation at the shock. In the LIA, the shock wave
is modeled as a planar discontinuity with no intrinsic scale. The
upstream flow, which consists of mean and fluctuating part, fully
determines the downstream flow via the Rankine-Hugoniot conditions at the
shock. In the LIA, turbulent field is decomposed into
individual Fourier modes. Each mode interacts independently with the
shock. Integration over all modes yields the full statistics of the
turbulent flow (cf.  Section~\ref{sec:lia}). 

In order to approximate the situation in CCSNe, we required the mean flow to have the vanishing Bernoulli parameter in the pre-shock region. We considered two types of upstream incident perturbations: the vorticity and entropy waves, both of which are advected with the mean flow. The vorticity mode is a solenoidal velocity field that has no pressure or density fluctuations, while the entropy mode represents density and temperature fluctuations with no associated pressure or velocity variations. Once the incident perturbations hit the shock, the downstream fluctuation field consists of vorticity, entropy, and acoustic waves (cf. Section~\ref{sec:lia}).

The compression factor $\cal C$ at the shock is the key quantity that
affects the flow through the shock. In particular, it determines
by how much the $x$-component of the wavenumber of incident waves increase as they cross the shock. Nuclear
dissociation leads to stronger compression: $\cal C$ increases from
$5.56$ to $10.15$ as nuclear dissociation parameter $\bar\epsilon$
increases from $0$ (inefficient nuclear dissociation) to $0.4$
(efficient nuclear dissociation). The compression factor $\cal C$ grows
fast with the upstream Mach number ${\cal M}_1$ until ${\cal M}_1 \sim\!
5$, after which it does not change much with further increasing of
${\cal M}_1$. We find most of the quantities that characterize the
downstream perturbation field have a similar dependence on ${\cal M}_1$
for ${\cal M}_1 \gtrsim 5$ (cf. Section~\ref{sec:results}). 

The critical angle $\psi_c$ separates two regions of the solution: $\psi_1 < \psi_c$ and $\psi_1 > \psi_c$, where $\psi_1$ is the incidence angle. The $\psi_1 < \psi_c$ region is called the propagative regime and it is characterized by acoustic waves in the post-shock flow, while $\psi_1 > \psi_c$ is called the non-propagative region, in which sound waves do not propagate. We investigated how $\psi_c$ depends on the free parameters of the mean flow: the upstream Mach number ${\cal M}_1$ and the efficiency of nuclear dissociation (cf. Section~\ref{sec:lia}). We find that $\psi_c$ depends weakly on both of these parameters. For our fiducial parameters, we get $\psi_c=69.2^\circ$ (cf. Section~\ref{sec:results}). 

We explored how individual vorticity and entropy waves affect the shock and the downstream flow (Section~\ref{sec:wave}). In particular, we analyzed the amplification of the kinetic energy of individual incident waves as they cross the shock. The amplification of kinetic energy does not change much with $\bar\epsilon$ and ${\cal M}_1$ for ${\cal M}_1\! \gtrsim\! 5$. On the other hand, it is highly sensitive to the relative phase between the entropy and vorticity waves: when they are in phase, we get the strongest amplification, while when they are out of phase, the kinetic energy does not amplify much. In fact, it may even decrease for some values of $\psi_1$. For example, for our fiducial parameters, ${\cal M}_1=5$ and $\bar\epsilon=5$, we get the amplification factors of $4.51$ and $0.90$ for in-phase and out-of-phase entropy-vorticity waves for $\psi_1=60^\circ$. The amplification for incident vorticity waves is roughly the average of these two regimes. For example, for the same values of $\bar\epsilon$, ${\cal M}_1$, and $\psi_1$, we get an amplification of $2.36$ (cf. Section~\ref{sec:wave}). 

For an incident field of turbulent fluctuations, we calculated the amplification of total turbulent kinetic energy. We find that the amplification is not sensitive to the nuclear dissociation parameter and the upstream Mach number beyond ${\cal M}_1 \gtrsim\! 5$. We again observe strong dependence on the phase difference between the incident vorticity and entropy waves. When they are in phase, the total kinetic energy increases by a factor of $\sim\!4$, while when they are out of phase, there is almost no amplification. When the phase is randomly distributed, the amplification is $\sim\!2$. When there is only incident vorticity wave perturbations, the amplification is again $\sim\!2$ (cf. Section~\ref{sec:turbulence}). 

When a turbulent eddy crosses the shock, it shrinks in size due to shock
compression. We find that for our fiducial values, the average linear
size of a turbulent eddy shrinks by a factor of $\simeq 2$ (cf.
Section~\ref{sec:turbulence}). This values does not change much with the
upstream Mach number ${\cal M}_1$ and the nuclear dissociation
parameter. This is somewhat disappointing news from the point of
producing explosion because smaller eddies are perhaps less likely to
become buoyant and help explosion \citep[e.g.,][]{couch:13b}.
However, this effect may be partially or completely offset by
the fact that perturbations in the nuclear-burning shells experience
significant radial stretching before reaching the shock
\citep{lai:00n,takahashi:14}.

When a turbulent field crosses the shock, the post-shock turbulence exerts additional pressure behind the shock. This reduces the critical neutrino luminosity necessary to drive the explosion \citep{mueller:15}. We find that, compared to the case with no post-shock turbulence, the critical luminosity decreases by a factor of $\simeq\!0.6 \langle {\cal M}_1'^2\rangle$, where $\langle {\cal M}_1'^2\rangle$ is the RMS turbulent Mach number in the pre-shock region. If the turbulent Mach number in convective shells is $\sim\! 0.1$, it may increase to $\langle {\cal M}_1'^2\rangle \sim\! 0.21$ during collapse prior to hitting the shock. This results in $\lesssim\!12\%$ reduction in the critical luminosity (cf. Section~\ref{sec:implications}).


\section*{Acknowledgements}

We thank Bernhard M\"uller for valuable discussions and comments. This
work is partially supported by ORAU and Social Policy grants at
Nazarbayev University and by the Sherman Fairchild Foundation. 

\appendix

\section{Nuclear Dissociation and the Shock Compression Factor}
\label{app:shock}

We choose our 1D shock parameters to approximate the CCSN shock by assuming vanishing Bernoulli parameter above the shock:
\begin{equation}
\label{eq:B}
B\equiv \frac{1}{2}v_1^2 + \frac{\gamma p_1}{(\gamma-1)\rho_1} - \frac{GM}{R} = 0,
\end{equation}
where $v_1$ represents the radial velocity immediately above the shock. We use $\gamma=4/3$ in all of our calculations. From Eq.~(\ref{eq:B}), we get
\begin{equation}
\frac{GM}{R} = \frac{1}{2}v_1^2 + \frac{\gamma p_1}{(\gamma-1)\rho_1} = \frac{1}{2}v_1^2 + \frac{c_\mathrm{s,1}^2}{\gamma-1},
\end{equation}
where $c_{\mathrm{s},1}$ is the speed of sound in the pre-shock region. Using the free-fall velocity, $v_\mathrm{ff}^2=2GM/R$, we rewrite the above equation as
\begin{equation}
v_\mathrm{ff}^2= v_1^2 + \frac{2 c_\mathrm{s,1}^2}{\gamma-1},
\end{equation}
Following \citet{fernandez:09a,fernandez:09b,radice:16a}, we parametrize $\epsilon$ as
\begin{equation}
\epsilon=\bar\epsilon \frac{1}{2}v_\mathrm{ff}^2
\end{equation}
where $\bar\epsilon$ is a dimensionless parameter that typically ranges from $0.2$ to $0.4$ \citep{fernandez:09a,fernandez:09b}. Using this definition, we can write the following expression for nuclear dissociation energy
\begin{equation}
\label{eq:eps}
\epsilon=\frac{1}{2}\bar\epsilon \left[v_1^2 + \frac{2 c_\mathrm{s,1}^2}{\gamma-1} \right]
\end{equation}
Our nuclear dissociation model affect the shock and the LIA
formalism by affecting the compression factor ${\cal C}$. The
latter is given by \citep{fernandez:09a}
\begin{equation}
\label{eq:c1}
{\mathcal C} = \frac{\gamma+1}{\gamma+\frac{1}{M_1^2} -
\sqrt{\left(1-\frac{1}{M_1^2}\right)^2+(\gamma^2-1)\frac{2\epsilon}{v_1^2}}},
\end{equation}
which depends on nuclear dissociated via the term $2\epsilon/v_1$. This, in turn, can be obtained
from Eq.~(\ref{eq:eps}):
\begin{equation}
\frac{2\epsilon}{v_1^2}=\bar\epsilon \left[1 + \frac{2}{\gamma-1}\frac{1}{{\cal M}_1} \right]
\end{equation}
Substituting this into Eq.~(\ref{eq:c1}), we get Eq.~(\ref{eq:c}) for $\cal C$:
\begin{equation}
{\mathcal C} = \frac{\gamma+1}{\gamma+\frac{1}{M_1^2} - \sqrt{\left(1-\frac{1}{M_1^2}\right)^2+(\gamma+1)\frac{(\gamma-1){\cal M}_1+2}{{\cal M}_1^2} \bar \epsilon}}.
\end{equation}
Note that this equations depends only on the upstream Mach number ${\cal
M}_1$ and the nuclear dissociation parameter $\bar \epsilon$. We fix our
units by setting $\rho_1=v_1=1$, which leaves us with only two free
parameters, ${\cal M}_1$ and $\bar \epsilon$, that fully specify the mean flow.

\section{The LIA formalism}
\label{app:lia}

For completeness, we present the LIA formalism in this section. Our presentation, including notation, closely follows that of \citet{mahesh:96}. The shock wave is modeled as planar discontinuity. The flow is decomposed into the mean and fluctuating parts. The latter is assumed to be weak so that the mean flow obeys the usual Rankine-Hugoniot conditions, while the perturbations obey the linearized version. The upstream flow completely determines the downstream flow and the shock dynamics. 

We start with the Rankine-Hugoniot conditions at the shock:
\begin{eqnarray}
\rho_1 v_1 &=& \rho_2 v_2, \\
p_1+\rho_1 v_1^2 &=& p_1+\rho_2 v_2^2, \\
\frac{1}{2} v_1^2 + \frac{\gamma p_1}{(\gamma-1)\rho_1} &=& \frac{1}{2}v_2^2 + \frac{\gamma p_2}{(\gamma-1)\rho_2}
\end{eqnarray}
where the subscript 1 and 2 denote pre- and post-shock
quantities. The quantities $\rho$, $p$, and $v$ are the
density, pressure, and the velocity of the flow. The stationary mean
flow is assumed to be in the positive $x$ direction and it is
characterized by its density $\bar\rho$, pressure $\bar p$, and
by the $x$-components of the velocity $U$. The
upstream perturbation field consists of entropy and vorticity waves
given by Eq.~(\ref{eq:u1})-(\ref{eq:T1}):
\begin{eqnarray}
\label{eq:u1a}
\frac{u_1'}{U_1} &=& lA_\upsilon e^{i\kappa(mx+ly-U_1mt)}, \\
\frac{\upsilon_1'}{U_1} &=& -mA_\upsilon e^{i\kappa(mx+ly-U_1mt)},\\
\label{eq:v1a}
\frac{\rho_1'}{\bar{\rho}_1} &=& A_e e^{i\kappa(mx+ly-U_1mt)}, \\
\label{eq:rho1a}
\frac{T_1'}{\overline{T}_1} &=& -\frac{\rho'}{\bar{\rho}_1},
\label{eq:T1a}
\end{eqnarray}
where $m=\cos\psi_1$, $l=\sin\psi_1$, and
$\psi_1$ is the angle between the $x$-axis and the direction of
propagation of the incident perturbation. $u_1'$ and $\upsilon_1'$ are
the $x$- and $y$-components of the velocity fluctuations, while
$A_\upsilon$ and $A_e$ are the amplitudes of the incident vorticity and
entropy waves. $\rho'$ and $T'$ are the density and temperature
perturbations. We ignore the acoustic waves in the upstream field.

When the upstream perturbations hit the
shock, the former responds by changing its position
and shape. In the LIA, for a perturbation of form
(\ref{eq:u1a})-(\ref{eq:T1a}), the shock surface deforms into a
sinusoidal wave propagating in the $y$-direction:
\begin{equation}
\label{eq:shock_lia0a}
\xi(y,t) = -\frac{L}{i\kappa m} e^{i\kappa (ly-U_1mt)}, \\
\end{equation}
where $\xi(y,t)$ is the $x$-coordinate of the shock position at ordinate $y$  and time $t$. $L$ is a quantity that characterizes the amplitude of the shock oscillations. The instantaneous velocity $\xi_t$ and inclination $\xi_y$ are given by
\begin{eqnarray}
\label{eq:shock_lia1a}
\xi_t(t,y) &=& U_1 L e^{i\kappa(ly-U_1mt)}, \\
\xi_y(t,y) &=& -\frac{l}{m} L e^{i\kappa(ly-U_1mt)}.
\label{eq:shock_lia2a}
\end{eqnarray}
The interaction of the vorticity and entropy waves with the shock generates a downstream perturbation field consisting of vorticity, entropy, and sound waves given by~\citep{mahesh:96,mahesh:97} 
\begin{eqnarray}
\label{eq:u2a}
&&\mkern-60mu\frac{u_2'}{U_1} = F e^{i\tilde{k}x}e^{i\kappa(ly-U_1mt)} + Ge^{ik(\mathcal{C}mx+ly-U_1 mt)}, \\ 
\label{eq:v2a}
&&\mkern-60mu\frac{\upsilon_2'}{U_1} = H e^{i\tilde{k}x}e^{i\kappa(ly-U_1mt)} + Ie^{i\kappa(\mathcal{C}mx+ly-U_1 mt)}, \\ 
\label{eq:p2a}
&&\mkern-60mu\frac{p_2'}{\bar{p}_2} = K e^{i\tilde{k}x}e^{ik(ly-U_1mt)} \\ 
&&\mkern-60mu\frac{\rho_2'}{\bar{\rho}_1} = \frac{K}{\gamma} e^{i\tilde{\kappa}x}e^{ik(ly-U_1mt)} + Qe^{i\kappa(\mathcal{C}mx+ly-U_1 mt)}, \\ 
\label{eq:T2a}
&&\mkern-60mu\frac{T_2'}{\overline{T}_1} = \frac{(\gamma-1)K}{\gamma} e^{i\tilde{\kappa}x}e^{i\kappa(ly-U_1mt)} -Qe^{i\kappa(\mathcal{C}mx+ly-U_1mt)}.
\end{eqnarray}
The schematic depiction of this process is given in
Fig.~\ref{fig:shocklia}. The coefficients $F$, $H$, and $K$ are the
amplitudes of the acoustic component, while coefficients $G$, $I$, and
$Q$ are associated with the entropy and vorticity components. The former
two components have the same wavenumber vector
$(m{\cal C} \kappa,l\kappa)$ and angular frequency $\kappa mU_1$. The
acoustic component has the same angular frequency but different
wavenumber $(\tilde{\kappa},l\kappa)$. In order to obtain the latter, we
write the wave equation for pressure in the post-shock region
\citep{mahesh:96}:
\begin{equation}
\label{eq:p}
p'_{tt}+2U_2 p'_{xt}-(c_\mathrm{s,2}^2-U_2^2)p'_{xx}-c_\mathrm{s}^2 p'_{yy} = 0,
\end{equation}
where $c_\mathrm{c,2}$ is the speed of sound. The solution in the post-shock region is required to have the same frequency and transverse wavenumber as the incoming perturbation. Thus, the general form of the solution of Eq.~(\ref{eq:p}) is
\begin{equation}
p'=F(x)e^{i\kappa(ly-mU_1t)}
\end{equation}
Assuming $F(x)\propto e^{\tilde\kappa x}$ and substituting this into Eq.~(\ref{eq:p}), we obtain a quadratic equation for $\tilde \kappa$
\begin{equation}
\left[\frac{c_\mathrm{s,2}^2}{U_1^2}-\frac{U_2^2}{U_1^2}\right] \tilde \kappa^2 + 2\kappa m \frac{U_2}{U_1}\tilde\kappa - \kappa^2 \left[m^2-l^2\frac{c_\mathrm{s,2}^2}{U_1^2}\right] = 0
\end{equation}
The discriminant of this equations is real if $\psi_1<\psi_c$ and complex if $\psi_1>\psi_c$, where the critical angle $\psi_c$ is given by Eq.~(\ref{eq:psic}):
\begin{equation}
\psi_c = \cot^{-1} \sqrt{\frac{c_{\mathrm{s},2}^2}{U_1^2} - \frac{U_2^2}{U_1^2}}.
\end{equation}
For $\psi_1<\psi_c$, $\tilde\kappa$ is real and is given by
\begin{equation}
\label{eq:kappa}
\frac{\tilde{\kappa}}{\kappa}=\frac{U_1}{U_2}\frac{M_2}{1-M^2_2}\bigg[-mM_2+l\; \sqrt[]{\frac{m^2}{l^2}- \frac{U^2_2}{U^2_1}\left(\frac{1}{M^2_2} -1\right) }\bigg],
\end{equation}
In this regime, the solution represents a simple sinusoidal planar sound wave. For $\psi_1>\psi_c$, $\tilde\kappa$ is complex, $\tilde\kappa=\tilde\kappa_r+i \tilde\kappa_i$:
\begin{eqnarray}
\label{eq:kappa_r}
\frac{\tilde{\kappa}_r}{\kappa}&=&- m\frac{U_1}{U_2}\frac{M^2_2}{1-M^2_2}, \\
\label{eq:kappa_i}
\frac{\tilde{\kappa}_i}{\kappa}&=& l\frac{U_1}{U_2}\frac{M_2}{1-M^2_2} \ \sqrt{\frac{U^2_2}{U^2_1}\left(\frac{1}{M^2_2}-1\right)-\frac{m^2}{l^2}},
\end{eqnarray}
This describes exponentially-damping planar sound wave. 

Our next task is to find the amplitudes of the post-shock solution. We start with the linearized Euler equations for the perturbation field \citep{mahesh:96}:
\begin{eqnarray} 
\label{eq:euler1}
u'_t+U_2u'_x &=& -\frac{1}{\bar\rho} p'_x, \\
\label{eq:euler2}
\upsilon'_t+U_2\upsilon'_x &=& -\frac{1}{\bar\rho} p'_y, 
\end{eqnarray} 
Substituting the acoustic part of the solution (\ref{eq:u2})-(\ref{eq:T2}) into the momentum equation in the $x$-direction (\ref{eq:euler1}), we get
\begin{equation}
U_1(-Fi\kappa mU_1)+U_2U_1Fi\tilde{\kappa}=-\frac{1}{\bar\rho_2}\bar p_2 Ki\tilde{\kappa}
\end{equation}
which can be solved for $F$:
\begin{equation}
F=\alpha K,
\end{equation}
where we introduced a new variable $\alpha$ for brevity:
\begin{equation}
\alpha = \frac{c_\mathrm{s,2}^2}{\gamma
U^2_1}\frac{\frac{\tilde{\kappa}}{\kappa}}{m-\frac{\tilde{k}}{kr}}.
\end{equation}
Analogously, the $y$-momentum equations yields
\begin{equation}
U_1(-Hi\kappa mU_1)+U_2U_1Hi\tilde{\kappa}=-\frac{1}{\bar{\rho_2}}P_2K i\kappa l,
\end{equation}
which we solve for $H$:
\begin{equation}
H=\beta K,
\end{equation}
where we introduced another variable $\beta$:
\begin{equation}
\beta = \frac{c_\mathrm{s,2}^2}{\gamma
U^2_1}\frac{l}{m-\frac{\tilde{\kappa}}{\kappa r}}.
\end{equation}
For vorticity waves, the velocity field has to be solenoidal:
\begin{equation}
U_1Gi\kappa mr + U_1Ii\kappa l=0,
\end{equation}
from which
\begin{equation}
I = - \frac{mr}{l}G.
\end{equation}
The Rankine-Hugoniot conditions at the shock shock yield the following equations for the downstream perturbation field:
\begin{eqnarray}
\frac{u^\prime_2 - \xi_t}{U_1}&=& B_1\frac{u^\prime_1- \xi_t}{U_1}+ B_2\frac{T^\prime_1}{\bar{T_1}},\\
\frac{\rho^\prime_2}{\bar{\rho_2}}&=&C_1\frac{u^\prime_1- \xi_t}{U_1}+C_2\frac{T^\prime_1}{\bar{T_1}}, \\
\frac{p^\prime_2}{\bar{P_2}}&=&D_1\frac{u^\prime_1- \xi_t}{U_1}+D_2\frac{T^\prime_1}{\bar{T_1}}, \\
\frac{\upsilon^\prime_2}{U_1}&=&\frac{\upsilon^\prime_1}{U_1}+E_1\xi_y,
\end{eqnarray}
where $A$, $B$, $C$, $D$, $E$ are function of upstream Mach number ${\cal M}_1$ and nuclear dissociating parameter $\bar \epsilon$ only. Substituting the downstream solution (\ref{eq:u2})-(\ref{eq:T2}) into these equations, we get a system of algebraic equations for the amplitudes of this solution  
\begin{eqnarray}
F+G-L&=&B_1(lA_\upsilon-L)-B_2A_e, \\
\frac{K}{\gamma}+Q&=&C_1(lA_\upsilon-L) - C_2A_e, \\ 
K &=& D_1(lA_\upsilon-L) - D_2A_e, \\ 
H+I &=& -mA_\upsilon - E_1\frac{l}{m}L.
\end{eqnarray}
We normalize the amplitudes of the solution (\ref{eq:u2})-(\ref{eq:T2}) with the amplitude of the incident vorticity wave $A_\upsilon$ (i.e., $\tilde F = F / A_\upsilon$, $\tilde L = L / A_\upsilon$, etc.). We rewrite the above system using these coefficients:
\begin{eqnarray}
\tilde{F} &=& \alpha \tilde{K}, \\ 
\tilde{H} &=& \beta \tilde{K}, \\
\tilde{I} &=& - \frac{mr}{l}\tilde{G}, \\ 
\tilde{F} + \tilde{G} - \tilde{L} &=& B_1(l-\tilde{L}) - B_2\frac{A_e}{A_\upsilon}, \\
\frac{\tilde{K}}{\gamma}+\tilde{Q}&=&C_1(l-\tilde{L})- C_2\frac{A_e}{A_\upsilon}, \\
\tilde{K}&=& D_1(l-\tilde{L})- D_2\frac{A_e}{A_\upsilon}, \\
\tilde{H}+\tilde{I}&=& - m - E_1\frac{l}{m}\tilde{L}. 
\end{eqnarray}
The solution of this system is 
\begin{eqnarray}
\tilde{L} &=& \frac{-m - \beta \left(D_1l-D_2\frac{A_e}{A_\upsilon}\right)}{E_1\frac{l}{m} - \beta D_1 - \frac{mr}{l}(1-B_1+\alpha D_1)} \\\nonumber &+& \frac{\frac{mr}{l}\left[-\alpha\left(D_1l-D_2\frac{A_e}{A_\upsilon}\right)+B_1l-B_2\frac{A_e}{A_\upsilon}\right]}{E_1\frac{l}{m} - \beta D_1 - \frac{mr}{l}(1-B_1+\alpha D_1)},
\end{eqnarray}
\begin{eqnarray}
\tilde{I}&=& - \frac{mr}{l} \bigg[ (1-B_1 + \alpha D_1)\tilde{L} \\\nonumber &-& \alpha \left(D_1l - D_2\frac{A_e}{A_\upsilon}\right) + B_1l - B_2\frac{A_e}{A_\upsilon} \bigg], 
\end{eqnarray}
\begin{eqnarray}
\tilde{G} &=& \tilde{L}(1-B_1 + \alpha D_1)\\\nonumber &-& \alpha \left(D_1l - D_2\frac{A_e}{A_\upsilon}\right) + B_1l - B_2\frac{A_e}{A_\upsilon},
\end{eqnarray}
\begin{eqnarray}
\tilde{K} &=& D_1(l- \tilde{L}) - D_2 \frac{A_e}{A_\upsilon}, \\
\tilde{F} &=& \alpha D_1 (l- \tilde{L}) - \alpha D_2 \frac{A_e}{A_\upsilon}, \\
\tilde{H} &=& \beta \bigg(D_1l - D_2 \frac{A_e}{A_\upsilon}\bigg) - \beta D_1 \tilde{L}, \\\nonumber 
\tilde{Q}&=&C_1(l-\tilde{L})- C_2\frac{A_e}{A_\upsilon} \\ && \quad - \frac{D_1}{\gamma}(l- \tilde{L}) + \frac{D_2}{\gamma} \frac{A_e}{A_\upsilon}.
\end{eqnarray}
We find that the solution depends on the upstream
Mach number ${\cal M}_1$ of the mean flow, the nuclear dissociation
parameter $\bar\epsilon$, and the ratio of the amplitudes of the
upstream entropy and vorticity waves $A_e/A_\upsilon$. Note that none of
the amplitude functions depend on the wavenumber $\kappa$ of the
incident waves, so the LIA solution is invariant with respect to the
spatial scale of the incoming perturbations.


\bsp
\label{lastpage}
\end{document}